%% file: LRPR_revise_tsp_2.tex
\documentclass[10pt]{IEEEtran} 

\usepackage{amsfonts}       
\usepackage{amsmath,amssymb,amsthm, bm, bbm,dsfont}
\usepackage{algorithm}
\usepackage{array}
\usepackage{mdwmath}
\usepackage{url}
\usepackage{multirow}

\usepackage{graphicx}
\usepackage{caption}
\usepackage{subcaption}

\usepackage{tikz}
\usepackage{algpseudocode}
\usepackage{pgfplots}
\newtheorem{theorem}{Theorem}[section]
\newtheorem{lemma}[theorem]{Lemma}
\newtheorem{corollary}[theorem]{Corollary}
\newtheorem{fact}[theorem]{Fact}
\newtheorem{definition}[theorem]{Definition}
\newtheorem{remark}[theorem]{Remark}

\newtheorem{sigmodel}[theorem]{Model}
\renewcommand\thetheorem{\arabic{section}.\arabic{theorem}}

\newcommand{\bproof}{ \begin{IEEEproof} }
\newcommand{\eproof}{ \end{IEEEproof} }
\newcommand{\beqno}{ \begin{equation*} }
\newcommand{\eeqno}{ \end{equation*} }
\newcommand{\beqa}{\begin{eqnarray*} }
\newcommand{\eeqa}{\end{eqnarray*} }
\newcommand{\beq}{ \begin{equation} }
\newcommand{\eeq}{ \end{equation} }

\newcommand{\tot}{q}
\renewcommand{\a}{\bm{a}}
\newcommand{\ta}{\bm{\tilde{a}}}
\newcommand{\x}{\bm{x}}
\renewcommand{\b}{\bm{b}}
\newcommand{\y}{\bm{y}}
\newcommand{\w}{\bm{w}}
\newcommand{\e}{\bm{e}}
\newcommand{\X}{\bm{X}}
\newcommand{\U}{{\bm{U}}}

\newcommand{\B}{\bm{B}}
\newcommand{\I}{\bm{I}}
\newcommand{\Y}{\bm{Y}}
\newcommand{\g}{\bm{g}}
\newcommand{\ghat}{\bm{\hat{g}}}
\newcommand{\xhat}{\bm{\hat{x}}}
\newcommand{\bhat}{\bm{\hat{b}}}
\newcommand{\Uhat}{{\bm{\hat{U}}}}

\newcommand{\n}{\mathcal{N}}
\newcommand{\E}{\mathbb{E}}
\newcommand{\one}{\mathds{1}}

\newcommand{\z}{\bm{z}}

\newcommand{\gd}{\bm{v}}
\newcommand{\ghatd}{\hat{\gd}}
\renewcommand{\ng}{\nu}
\newcommand{\hatng}{\hat{\ng}}

\newcommand{\iidsim}{\stackrel{\mathrm{iid}}{\thicksim }}
\newcommand{\indepsim}{\stackrel{\mathrm{indep}}{\thicksim }}
\newcommand{\SE}{\mathrm{SE}}
\newcommand{\dist}{\mathrm{dist}}
\newcommand{\tta}{\ta^{\text{trunc}}}

\newcommand{\A}{\bm{A}}
\newcommand{\C}{\bm{C}}
\newcommand{\Chat}{\bm{\hat{C}}}

\newcommand{\M}{\bm{M}}
\newcommand{\F}{\bm{F}}

\newcommand{\Span}{\mathrm{range}}
\newcommand{\trace}{\mathrm{trace}}

\newcommand{\cM}{\hat{\bm{D}}}  
\newcommand{\cA}{\bm{D}}    
\newcommand{\cH}{\bm{H}}
\newcommand{\bE}{\bm{E}}
\newcommand{\bF}{\bm{F}}
\newcommand{\total}{\mathrm{tot}}
\newcommand{\Lamk}{\bm\Lambda_k}
\newcommand{\Lambar}{\bar{\bm\Lambda}}
\newcommand{\lambar}{\bar{\lambda}}
\newcommand{\evdeq}{\overset{\mathrm{EVD}}=} 
\newcommand{\lammax}{\bar\lambda_{\max}} 
\newcommand{\lammin}{\bar\lambda_{\min}}

\newcommand{\tildm}{{\tilde{m}}}

\newcommand{\N}{\mathcal{N}}

\newcommand{\W}{\bm{W}}
\renewcommand{\Re}{\mathbb{R}}
\newcommand{\rhat}{\hat{r}}
\newcommand{\mat}{{\mathrm{mat}}}

\newcommand{\bi}{\begin{itemize}}
\newcommand{\ei}{\end{itemize}}
\newcommand{\Xhat}{\hat{\bm{X}}}

\begin{document}
%

\title{Low Rank Phase Retrieval}
\author{Namrata Vaswani, Seyedehsara Nayer, Yonina C. Eldar

\thanks{N. Vaswani and S. Nayer are with the Iowa State University, Ames, IA, USA. Y. C. Eldar is with Technion, Haifa, Israel.  Email: namrata@iastate.edu.
An early version of the initialization idea developed in this work was presented at the IEEE Statistical Signal Processing Workshop 2016 \cite{ssp_pr}. A short version of this work will be presented at ICASSP 2017 \cite{lrpr_icassp}.
}
}

\maketitle

\begin{abstract} 
We develop two iterative algorithms for solving the low rank phase retrieval (LRPR) problem. LRPR refers to recovering a low-rank matrix $\X$ from magnitude-only (phaseless) measurements of random linear projections of its columns. Both methods consist of a spectral initialization step followed by an iterative algorithm to maximize the observed data likelihood. We obtain sample complexity bounds for our proposed initialization approach to provide a good approximation of the true $\X$. When the rank is low enough, these bounds are significantly lower than what existing single vector phase retrieval algorithms need. Via extensive experiments, we show that the same is also true for the proposed complete algorithms.
\end{abstract}


\section{Introduction}
\label{intro}
In recent years there has been a large amount of work on the phase retrieval (PR) problem and on its generalization. The original PR problem involves recovering a length-$n$ signal $\x$ from the magnitudes of its discrete Fourier transform (DFT) coefficients. 
Generalized PR replaces the DFT by inner products with any set of measurement vectors, $\a_i$. Thus, the goal is to recover $\x$ from $|\a_i{}'\x|^2$, $i=1,2, \dots, m$. These magnitude-only measurements are referred to as {\em phaseless measurements}. 
%
PR is a classical problem that occurs in many applications such as X-ray crystallography, astronomy, and ptychography because the phase information is either difficult or impossible to obtain \cite{SECCMS15}. Algorithms for solving it have existed since the work of Gerchberg and Saxton and Fineup \cite{ger_saxton,fineup}.
In recent years, there has been much renewed interest in PR, e.g., \cite{candes2013phase,candes_phaselift,eldar:14:prs,pr_altmin,wf,twf,SECCMS15,bauschke2002phase,bauschke2005new, mukherjee2012iterative,waldspurger2015phase,fogel2013phase,chouzenoux2013block,repetti2014nonconvex} and  in sparse PR, e.g., \cite{jaganathan2012recovery,shechtman2013gespar,szameit:12:sbs}.


One popular class of approaches, pioneered in Cand\`{e}s et al. \cite{candes2013phase,candes_phaselift}, solves PR by recovering the rank one matrix $\bm{Z}:=\x \x'$ from $\y_i:=|\a_i{}'\x|^2 = \mathrm{trace}(\a_i \a_i{}' \bm{Z})$ via a semi-definite relaxation.
This can provably recover $\x$ (up to a global phase uncertainty) using only $m=c n$ independent identically distributed (iid) Gaussian phaseless measurements. However because of the `lifting', its computational and storage complexity depends on $n^2$ instead of on $n$ (it needs to recover an $n\times n$ matrix instead of an $n$-length vector). {\em  Here, and throughout the paper, the letter $c$ is re-used to denote different numerical constants each time it is used.}

In more recent works, non-convex methods, that do not lift the problem to higher dimensions, have been explored along with provable guarantees \cite{pr_altmin,wf,twf}. An alternating minimization (AltMin) technique with spectral initialization, AltMinPhase, was developed and analyzed in \cite{pr_altmin}. The AltMin step of this approach is essentially the same as the old Gerchberg-Saxton algorithm \cite{ger_saxton}. A gradient descent method with spectral initialization, called Wirtinger Flow (WF), was studied in \cite{wf}.
In \cite{twf}, truncated WF (TWF), which introduced a truncation technique to further improve WF performance, was developed.
It was shown that TWF recovers $\x$ from only $c n$ iid Gaussian phaseless measurements, while the number of iterations needed for getting an error of  order $\epsilon$ is $c \log (1/\epsilon)$ (converges geometrically). AltMinPhase and WF require more measurements, $cn \log^3 n$ and $cn \log n$, respectively. WF also has a slower convergence rate. 
Two recent modifications of TWF \cite{rwf,taf} have the same order complexities but improved empirical performance.


%

{\bf Problem Setting. }
In this work, instead of a single vector $\x$, we consider a set of $q$ vectors, $\x_1,\x_2, \dots, \x_q$, such that the $n \times q$ matrix,
\[
\X:=[\x_1,\x_2, \dots, \x_\tot],
\]
has rank $r \ll \min(n,\tot)$.
For each column $\x_k$ of $\X$, we observe a set of $m$ measurements of the form
\begin{eqnarray}
\y_{i,k}:= |\a_{i,k}{}'\x_k|^2, \ i = 1, 2, \dots m, \ k=1,2, \dots, \tot.
\label{exact_mod}
\end{eqnarray}
The measurement vectors, $\a_{i,k}$, are mutually independent.
Our goal is to recover the matrix $\X$ from these $m q$ phaseless measurements $\y_{i,k}$.  Since we have magnitude-only measurements of each column $\x_k$, we can only hope to recover each column $\x_k$ up to a global phase ambiguity. We refer to the above problem as {\em low rank phase retrieval (LRPR)}.

In some applications, the goal may be to only recover the span of the columns of $\X$, $\Span(\X)$.  This would be the case, for example, if one is interested in only seeing the principal directions of variation of the dataset, and not in recovering the dataset itself. 
We refer to this easier problem as {\em phaseless PCA (principal component analysis)}.

A motivating application for LRPR is dynamic astronomical imaging such as solar imaging where the sun's surface properties gradually change over time \cite{butala}.
The changes are usually due to a much smaller number of factors, $r$, than the size of the image, $n$, or the total number of images, $\tot$.
If the images are arranged as 1D vectors $\x_k$, then the resulting matrix is approximately low rank.
%
As another potential application, consider a Fourier ptychography imaging system that captures a dynamic scene exhibiting a temporal evolution; this is often the case when observing live biological specimens {in vitro}. Suppose the scene resolution is $n$ and the total number of captured frames is $q$. If the dynamics is approximated to be linear and slow changing, then the matrix formed by stacking the frames next to each other can be modeled as a rank-$r$ matrix, where $r \ll \min(q,n)$.  Similar applications involving a sequence of gradually changing images also occur in X-ray and sub-diffraction imaging systems. Moreover, if we are only interested in identifying the principal directions of variation of the image sequences, then the problem becomes that of phaseless PCA. 
PCA is often the first step for classification, clustering, modeling, or other exploratory data analysis. 






{\bf Contributions. } This work has two contributions. We propose iterative algorithms for solving the LRPR problem described above. Our solution approach relies on the fact that a rank $r$ matrix $\X$ can be expressed (non-uniquely) as $\X = \U \B$ where $\U$ is an $n \times r$ matrix with mutually orthonormal columns. Its first step consists of a spectral initialization step, motivated by TWF, for first initializing $\U$, and then, the columns of $\B$. The remainder of the algorithm is developed in one of two ways:  using a projected gradient descent strategy to modify the TWF iterates (LRPR1); or an AltMin algorithm, motivated by AltMinPhase, that directly exploits the decomposition $\X = \U \B$ (LRPR2). Via extensive experiments, we demonstrate that both LRPR1 and LRPR2 have better sample complexity than TWF; with LRPR2 being the best. 
Moreover, when enough measurements are available for TWF to work,  we show that the LRPR initialization can also be used to speed up basic TWF for solving LRPR.

Our second, and most important, contribution is a sample complexity bound for the proposed initialization to get within an $\varepsilon$ ball of the true $\X$.
Our results show that, if the goal is to only initialize $\U$ with subspace recovery error below a fixed level, say $\varepsilon=1/4$, then a total of $mq = c n r^2 /\varepsilon^2 = 16 cnr^2$ iid Gaussian measurements suffice with high probability (whp). When $r$ is small, $nr^2$ is only slightly larger than $nr$ which is the minimum required by any technique to recover the span of $\U$. If the goal is to also initialize the $\x_k$'s with normalized error below say $\varepsilon=1/4$, then we need more measurements, but still significantly fewer than TWF. For example, if $r \le c \log n$ and $q \ge c n$, then, only $16 c\sqrt{n}$ measurements per column are required.
We note that our guarantees assume that a different set of measurements is used for initializing $\U$ and $\B$ (see Model \ref{meas_mod}).

As seen in many earlier works, e.g., AltMinPhase \cite{pr_altmin}, resampled WF \cite[Algorithm 2 and Theorem 5.1]{wf} or TWF \cite{twf}, the sample complexity of the entire algorithm is equal to or smaller than that of the initialization step for a fixed error level\footnote{For AltMinPhase, the initialization sample complexity (for achieving a given fixed error) is $c n \log^3 n$ while it is only $c n \log n$ per iteration for the rest of the algorithm. For resampled WF, it is $c n \log^2 n$ for initialization and $c n \log n$ for the rest of the algorithm, while for TWF, it is $c n$ both for the initialization and for the complete algorithm.}. This is why initialization guarantees are important.


Our problem setting assumes {\em a different (mutually independent) set of measurement vectors is used for imaging each column $\x_k$}. This is critical for guaranteeing the improved sample complexity of our solution approach over single-vector PR methods because this is what ensures that the $mq$ matrices $\y_{i,k} \a_{i,k} \a_{i,k}{}'$ are all mutually independent conditioned on $\X$. Hence, we can exploit averaging over $mq$ such matrices when estimating $\U$.
If $\a_{i,k} = \a_{i,1}$ (same $\a_i$'s are used), then this benefit disappears since only $m$ of the above matrices are mutually independent. We demonstrate this in Table \ref{real_gauss} (last column) in Sec. \ref{sims}. We discuss the practical implications of our setting in Sec. \ref{indep_meas}.

Two other works that also generalize WF \cite{wf}, but to solve a completely different problem include \cite{procrustes,zheng_lafferty}. These study the problem of recovering a rank $r$ matrix $\M$ from measurements of the form $\z_i = \mathrm{trace}(\bm{A}_i{}'\M)$. This is the low rank matrix sensing problem studied in \cite{lowrank_altmin} and a lot of earlier and later works.
In our problem, if we use the {\em same} $\a_i$'s for all columns $\x_k$, and define $\z_i:= \sum_k (\a_i{}'\x_k)^2$ and $\M := \sum_k \x_k \x_k{}'$, then we could use the strategy of \cite{procrustes,zheng_lafferty} or, in fact, any low-rank matrix sensing technique, e.g., AltMinSense from \cite{lowrank_altmin}, to recover $\M$ from the $\z_i$'s; followed by recovering $\Span(\U)$ as its column space. However, for the reasons explained above, use of same measurement vectors will not yield any advantage over single vector PR.
When using different $\a_{i,k}$'s, none of these methods are applicable.

\newcommand{\polylog}{\mathrm{polylog}}

{\bf Notation. }
The notation $\a_{i,k} \stackrel{\text{iid}}{\sim} \n(\bm\mu, \bm\Sigma)$ means that the vectors $\a_{i,k}$ are iid real Gaussian vectors with mean $\bm\mu$ and covariance matrix $\bm\Sigma$; and $\b_k \indepsim \n(\bm\mu_k,\bm\Sigma_k)$ means that the $\b_k$'s are mutually independent and $\b_k$ is generated from $\n(\bm\mu_k,\bm\Sigma_k)$.
We use $'$ to denote matrix or vector conjugate transpose, and $\|.\|_p$ to denote the $l_p$ norm of a vector or the induced $l_p$ norm of a matrix. When the subscript $p$ is missing, i.e., when we just write $\|.\|$, it denotes the $l_2$ norm of a vector or the induced $l_2$ norm of a matrix. We use $\I$ to denote the identity matrix.
The notation $\mathbbm{1}_{\zeta}$ is the indicator function for statement $\zeta$, i.e., $\mathbbm{1}_{\zeta}=1$ if $\zeta$ is true and $\mathbbm{1}_{\zeta}=0$ otherwise.
For a vector $\z$, $|\z|$, $\sqrt{\z}$ and $\mathrm{phase}(\z)$ compute the {\em element-wise} magnitude, square-root, and phase
of each entry of $\z$, and $\mathrm{diag}(\z)$ creates a diagonal matrix with entries from $\z$.




{\bf Paper Organization. }  In Sec. \ref{algos}, we develop the proposed LRPR initialization approach (LRPR-init). We obtain sample complexity bounds for it in Sec. \ref{guarantees}. In Sec. \ref{algo_iters}, we explain how LRPR-init can be used to develop iterative algorithms for LRPR that are either faster than basic TWF (LRPR+TWF) or need a smaller $m$ to work (LRPR1 and LRPR2). Numerical experiments backing our claims are shown in Sec. \ref{sims}.
We prove our results from Sec. \ref{guarantees} in Sec. \ref{outline}, and conclude in Sec.~\ref{conclusions}.%

The algorithms proposed in this work are applicable for both real and complex measurements. Experiments are shown for both cases too. Moreover, as shown in our experiments, our algorithms also apply to noisy measurements. However, for simplicity, we state and prove our guarantees only for the real Gaussian measurements' case. Their extension to complex Gaussian measurements is straightforward.


\section{Low Rank PR (LRPR) Initialization} 
\label{algos}

Our goal is to recover an $n \times q$ low rank matrix $\X$ from phaseless measurements of linear projections of each of its columns, i.e, from $\y_{i,k}:=(\a_{i,k}{}'\x_k)^2$, $i=1,2,\dots,m$, and $k=1,2,\dots, q$.
In this section, we develop an approach to obtain an initial estimate of $\X$ that relies on the fact that a rank $r$ matrix $\X$ can be expressed (non-uniquely\footnote{We can rewrite $\X$ as $\X = (\U \bm{R}) (\bm{R}' \B)$ for any rotation matrix $\bm{R}$.}) as $\X = \U \B$ where $\U$ is an $n \times r$ matrix with mutually orthonormal columns and $\B = [\b_1, \b_2, \dots \b_\tot]$ is an $r \times \tot$ matrix. 
The proposed initialization approach first computes an estimate of $\Span(\U)$, i.e., it returns $\Uhat$ that may be very different from $\U$ in Frobenius norm, but their spans are close, i.e., the subspace error, $\SE(\Uhat, \U)$ is small. Here,
\[
\SE(\Uhat,\U):= \| (\I - \Uhat \Uhat{}') \U \|
\]
quantifies the subspace error (principal angle) between the range spaces of two matrices $\Uhat,\U$ with mutually orthonormal columns.
%
Using $\Uhat$, we find estimates $\bhat_k$ so that $\dist(\Uhat \bhat_k , \x_k)$ is small.
Here,
\[
\dist(\z_1, \z_2) :=  \min_{\phi \in [0,2\pi]} \| \z_1 - e^{\sqrt{-1} \phi} \z_2\|
\]
quantifies the phase-invariant distance\footnote{When $\z_1$ and $\z_2$ are both real, the phase is only $+1$ or $-1$, and so, $\dist(\z_1, \z_2)= \min(\|\z_1-\z_2\|,\|\z_1+\z_2\|)$.} between two  complex vectors $\z_1, \z_2$ \cite{wf,twf}.
The $\b_k$'s are initialized by estimating%
\[
\g_k:=\Uhat'\x_k=\Uhat'\U \b_k
\]
for each $k$, and setting $\bhat_k = \ghat_k$. Because $\Uhat$ can be arbitrarily rotated w.r.t. $\U$, this approach may not give accurate estimates of the individual $\b_k$'s, i.e.,  $\dist(\b_k,\bhat_k)$ may not be small.

\subsection{LRPR-init: Spectral initialization for LRPR}
LRPR-init is a two step approach. We first initialize $\U$ using a truncated spectral initialization idea \cite{twf}. For this, define
\[
\Y_{U,0}:=\frac{1}{m q} \sum_{i=1}^m \sum_{k=1}^q \y_{i,k} \a_{i,k} \a_{i,k}{}'.
\]
Let
$
\frac{1}{q} \sum_{k=1}^q \x_k \x_k{}'  \evdeq \U \Lambar \U'
$
denote the reduced eigenvalue decomposition (EVD) of $\X \X' / q$. Thus, $\U$ is an $n \times r$ matrix with orthonormal columns and $\Lambar$ is an $r \times r$ diagonal matrix.
It is not hard to see that \cite[Lemma A.1]{wf},
\beq
\E[\y_{i,k} \a_{i,k} \a_{i,k}{}'] =  2 \x_k \x_k{}'  + \|\x_k\|^2 \I. \label{LemA1_wf}
\eeq
and, therefore,
\[
\E[\Y_{U,0}] = 2\U \Lambar \U' + \trace(\Lambar)\I.
\] 
Clearly, the subspace spanned by the top $r$ eigenvectors of this matrix is equal to $\Span(\U)$ and the gap between its $r$-th and $(r+1)$-th eigenvalue is $2\lambda_{\min}(\Lambar)$.
If $m$ and $q$ are large enough, then, one can use an appropriate law of large numbers' result to argue that $\Y_{U,0}$ will be close to its expected value whp.
By the $\sin \theta$ theorem \cite{davis_kahan}, as long as $2\lambda_{\min}(\Lambar)$ is large compared to $\| \Y_{U,0} - \E[\Y_{U,0}]\|$, the same will also be true for the span of the top $r$ eigenvectors of $\Y_{U,0}$.

However, as explained in \cite{twf}, because $\y_{i,k} \a_{i,k} \a_{i,k}{}'$ can be written as $\w \w'$ with $\w$ a heavy-tailed random vector, more samples will be needed for the law of large numbers to take effect than if $\w$ were not heavy-tailed. 
To remedy this situation, we use the truncation idea suggested in \cite{twf} and compute $\Uhat$ as the top $r$ eigenvectors of
\beq \label{def_YU}
\Y_U:=\frac{1}{m \tot} \sum_i \sum_k  \y_{i,k} \a_{i,k} \a_{i,k}{}' \mathbbm{1}_{ \{ y_{i,k}\leq  9 \frac{\sum_i y_{i,k}}{m} \} } .
\eeq
The idea of truncation is to average only over those $(i,k)$'s for which $\y_{i,k}$ is not too far from its empirical mean. 

Next we consider initialization of the $\b_k$'s. Define the matrix
\beq
\bm{M}_k:=\frac{1}{m} \sum_i \y_{i,k} \a_{i,k} \a_{i,k}{}'.
\label{def_Mk}
\eeq
Suppose that $\Uhat$ is independent of the $\bm{M}_k$'s. 
Then, from \eqref{LemA1_wf}, conditioned on $\Uhat$, 
\[
\E[\Uhat{}' \bm{M}_k  \Uhat ] =
\Uhat' ( 2 \x_k \x_k{}'  + \|\x_k\|^2 \I ) \Uhat =  2 \g_k \g_k{}' + \|\x_k\|^2 \I.  
\]
The top eigenvector of this expectation is proportional to $\g_k$ and the gap between its first and second eigenvalues is $2\|\g_k\|^2 = 2\|\Uhat' \U \b_k\|^2$.
Thus, as long as $\Uhat$ is a good estimate of $\U$ (in terms of $\SE$), the eigen-gap will be close to $2\|\b_k\|^2$.
Therefore, we can argue that the normalized top eigenvector of $\Uhat{}' \bm{M}_k  \Uhat$, denoted $\ghatd_k$, will be a good estimate of $\gd_k:= \g_k/\|\g_k\|$. Using this idea, we initialize the $\x_k$'s as $\xhat_k  = \Uhat \ghatd_k  \hatng_k$ where $\hatng_k = \sqrt{\sum_i \y_{i,k} / m}$ is an estimate of $\ng_k:=\|\g_k\|$.
We do not use truncation here because $\g_k$ is an $r$ length vector, with $ r \ll n$, and we need to use many more than $r$ measurements for accurate recovery.

The complete approach, LRPR-init, is summarized in Algorithm \ref{algo_init}. Note that this uses the same set of measurements to recover $\U$ and $\b_k$'s. But, as seen from our numerical experiments, it still works well in practice. For our analysis in Sec.~\ref{guarantees}, we assume that a new set of measurements is available for computing $\ghat_k$, and thus $\bm{M}_k$ is independent of $\Uhat$.

Algorithm \ref{algo_init} also estimates the rank $r$ automatically by looking for the maximum gap between consecutive eigenvalues of $\Y_U$. As we explain in Sec. \ref{unknown_r}, under a simple assumption on the eigenvalues of $\Lambar$, this returns the correct rank whp. 


\begin{algorithm}[h]
\caption{\small{Low Rank PR Initialization (LRPR-init)}} 
\label{algo_init}
Set $\hat{r} = \arg\max_j (\lambda_j(\Y_U)-\lambda_{j+1}(\Y_U) )$ with $\Y_U$  defined in \eqref{def_YU}.
\begin{enumerate}
\item Compute $\Uhat$ as top $\hat{r}$ eigenvectors of $\Y_U$.

\item For each $k=1,2, \dots, \tot$,
\begin{enumerate}
\item compute $\ghatd_k$ as the top eigenvector of $\Uhat{}' \frac{1}{m} \sum_i \y_{i,k} \a_{i,k} \a_{i,k}{}'  \Uhat$.

\item compute $\hatng_k:= \sqrt{ \frac{1}{m} \sum_{i}\y_{i,k} }$; set $\bhat_k = \ghat_k= \ghatd_k \hatng_k$
\end{enumerate}
\end{enumerate}
Output $\Uhat$ and $\xhat_k:= \Uhat \bhat_k$ for all $k=1,2,\dots,q$. 
\end{algorithm}

\subsection{Projected-TWF initialization}
Another way to obtain an initial estimate of the low rank matrix $\X$ would be to project the matrix formed by the TWF initialization for each column $\x_k$ onto the space of rank $r$ matrices. This is summarized in Algorithm \ref{twfproj_init_algo}. However, as we show in Sec. \ref{sims}, Tables \ref{real_gauss} and \ref{complex_gauss}, this approach performs much worse than LRPR-init. The reason is that it does not simultaneously exploit averaging of the matrices $ \y_{i,k} \a_{i,k} \a_{i,k}{}'$ over both $i$ and $k$.

\begin{algorithm}
\caption{Projected-TWF initialization (TWFproj-init)}
\label{twfproj_init_algo}
\begin{enumerate}
\item For each $k=1,2,\dots,q$, set $\xhat_k^0$ as the top eigenvector of $\frac{1}{m}\sum_{i=1}^m \y_{i,k} \a_{i,k} \a_{i,k}{}'\mathbbm{1}_{ \{ y_{i,k}\leq  9 \frac{\sum_i y_{i,k}}{m} \} }$ scaled by $\sqrt{\sum_{i=1}^m \y_{i,k} / m }$; create $\Xhat^{0,TWF}$
\item Project $\Xhat^{0,TWF}$ onto the space of rank $r$ matrices to get $\Xhat^0$.
\end{enumerate}
\end{algorithm}

\section{Sample Complexity Bounds for LRPR-init} \label{guarantees}
In this section, we obtain sample complexity bounds for getting a provably accurate initial estimate of both $\U$ and of the $\x_k$'s whp. For simplicity, our results assume iid real Gaussian measurement vectors, $\a_{i,k}$. As will be evident from the proofs, the extension to complex Gaussian vectors is straightforward.
In Sec. \ref{results_det}, we provide a guarantee for the case when $\X$ is a deterministic unknown matrix with known rank $r$. These hold whp over measurement vectors $\a_{i,k}$.
In Sec. \ref{results_gauss}, we give results for the case of $\X$ being random with known rank $r$. These hold whp both over matrices $\X$ generated from the assumed probability distribution and over measurement vectors $\a_{i,k}$. In Sec. \ref{unknown_r}, we show how we can extend both sets of results to the unknown rank case.

The proof of our results consists of two parts. We first bound the subspace recovery error $\SE(\Uhat,\U)$. Next, we use this to bound the error in estimating the $\x_k$'s, $\dist(\xhat_k, \x_k)$. To do this, we show that, if $\Uhat$ is a given matrix with $\SE(\Uhat,\U)$ small enough, and if the measurement vectors and the measurements that are used to estimate the $\b_k$'s are independent of $\Uhat$, then, whp, $\dist^2(\xhat_k, \x_k)$ can be shown to be bounded by $c \varepsilon \|\x_k\|^2$ for any chosen $\varepsilon$.
To ensure that the independence assumption holds, we use a standard trick developed in many earlier works, e.g., \cite{pr_altmin}. We analyze a ``partitioned" version of Algorithm \ref{algo_init}. Denote the total number of measurements by $m_{\total}$. We partition these into two disjoint sets of size $m$ and $\tilde{m}$ respectively; we use the first set for estimating $\U$ and the second set for estimating the $\b_k$'s.
Denote the first set of measurements and measurement vectors by $\y_{i,k}$ and $\a_{i,k}$ respectively. Denote the second set by $\y_{i,k}^{new}$ and $\a_{i}^{new}$ respectively.
Since the different $\b_k$'s are recovered independently, for the second set, we can use the same measurement vectors, $\a_i^{new}$, for all the $\x_k$'s.
Thus, we have the following setting.

\begin{sigmodel}[Measurement model]\label{meas_mod} \em
For each $\x_k$, 
\begin{itemize}
\item we observe $\y_{i,k}:= (\a_{i,k}{}'\x_k)^2$ where  $\a_{i,k} \iidsim \n(0, \I)$, for $i=1,2,\dots,m$; 
and
\item we observe $\y_{i,k}^{new}:= (\a_{i}^{new}{}'\x_k)^2$ where $\a_{i}^{new} \iidsim \n(0, \I)$, for $i=1,2,\dots,\tilde{m}$.
\item  The sets of vectors $\{\a_{i}^{new}, i=1,2,\dots,\tilde{m} \}$ and $\{\a_{i,k}, {i=1,2,\dots,m, \ k=1,2,\dots,q} \}$ are mutually independent.%
\end{itemize}
Thus we have a total of $m_{\total} = m + \tilde{m}$ measurements per vector $\x_k$.
\end{sigmodel}

\begin{algorithm}[t!]
\caption{\small{LRPR-init-theoretical: initialization with partitioned measurements.}} 
 \label{algo_init_disjoint}

{\bf Known $r$: } Set $\rhat = r$.

{\bf Unknown $r$: } Set $\hat{r} = \arg\max_j (\lambda_j(\Y_U)-\lambda_{j+1}(\Y_U) )$ where $\Y_U$ is defined in \eqref{def_YU}.

\begin{enumerate}
\item Compute $\Uhat$ as top $\rhat$ eigenvectors of $\Y_U$ defined in \eqref{def_YU}.

\item For each $k=1,2, \dots, q$,
\begin{enumerate}
\item
compute $\ghatd_k$ as the top eigenvector of 
\begin{equation} \label{def_Yb}
\Y_{b,k}:= \Uhat{}'\left(\frac{1}{\tilde{m}} \sum_{i=1}^{\tilde{m}} \y_{i,k}^{new} \a_{i}^{new} \a_{i}^{new}{}'  \right) \Uhat,
\end{equation}

\item compute $\hatng_k= \sqrt{ \frac{1}{\tilde{m}} \sum_{i} \y_{i,k}^{new} }$; set $\bhat_k=\ghat_k =\ghatd_k \hatng_k$.
\end{enumerate}

\end{enumerate}
Output $\Uhat$ and $\xhat_k:= \Uhat \bhat_k$ for all $k=1,2,\dots,q$.
\end{algorithm}

With measurements taken as above, we study Algorithm \ref{algo_init_disjoint}.

\subsection{Main Results for Deterministic $\X$ - Known rank case} \label{results_det} 


Let
\beq \label{def_Lambar}
\frac{1}{q}\X \X' \evdeq \U \Lambar \U',  
\eeq
and $\X = \U \B$.
Thus, $\Lambar = \frac{1}{q} \sum_k \b_k \b_k{}'$.
Let $\lammax$ and $\lammin$ denote the maximum and minimum eigenvalues of $\Lambar$.
Define
\beq \label{def_rho}
\rho:=   \frac{\max_{k} \|\x_k\|^2}{\frac{1}{q} \sum_{k} \|\x_k\|^2}, \text { and } \kappa:= \frac{\lammax}{\lammin}.
\eeq
Thus, $\kappa$ is the condition number of $\X \X'$. Using $\rho$, we can bound $\max_k \|\b_k\|^2=\max_k \|\x_k\|^2$ in terms of $\lammax$ as\footnote{This follows because $\frac{1}{q} \sum_k \|\x_k\|^2=  \trace(\frac{1}{q} \sum_k \x_k \x_k{}') = \trace(\frac{1}{q}\X \X') = \trace(\Lambar) = \sum_{j=1}^r \lambda_j(\Lambar)$.}
\[
\max_k \|\b_k\|^2=\max_k \|\x_k\|^2 = \rho \sum_{j=1}^r \lambda_j(\Lambar) \le r \rho \lammax.
\]
We then have the following result.

\begin{theorem}[Deterministic $\X$] \label{mainthm}
Consider an unknown deterministic rank $r$ matrix $\X$. Assume that the measurements of its columns are generated according to Model \ref{meas_mod}.
Consider the output of Algorithm \ref{algo_init_disjoint} (known $r$ case). Suppose that $r \le c n^{1/5}$.
For an $\varepsilon < 1$, if
\begin{align*}
& \tilde{m} \ge \frac{ c\sqrt{n}}{\varepsilon^2}, \
 m \ge \frac{c \kappa^2  \cdot r^4 \log n  (\log \tilde{m})^2}{\varepsilon^2}, \\
& mq \ge \frac{c  \rho^2 \kappa^2  \cdot n  r^4 (\log \tilde{m})^2 }{\varepsilon^2},
\end{align*}
then, with probability at least $1 -  4\exp(-cn) - \frac{32q}{n^4}$,  
\begin{enumerate}
\item
\[
\SE(\Uhat,\U) 
\le   \frac{c \varepsilon}{r \log \tilde{m}};
\]
\item
for all $k=1,2,\dots,q$,  $\dist( \x_k , \xhat_k )^2 \le c  \varepsilon \|\x_k\|^2$, and so
\[
\mathrm{NormErr}(\X,\hat\X):= \frac{\sum_{k=1}^q \dist( \x_k , \xhat_k )^2 }{ \sum_{k=1}^q \|\x_k\|^2 } \le c \varepsilon.
\]
\end{enumerate}
Furthermore, if $q \le c n^2$, then the above event holds with probability at least $1 - c /n^2$.
\end{theorem}
\begin{proof}
The proof is given in Section \ref{outline}. 
\end{proof}
Notice that our lower bounds depend on $\kappa^2$ where $\kappa$ is the condition number of $\X \X'$. This is pretty typical, e.g., it is also the case in \cite{lowrank_altmin,procrustes,zheng_lafferty} and many other works. It may be possible to remove this dependence by borrowing ideas from \cite{lowrank_altmin_no_kappa}. A second point to note is that the probability of the good event depends inversely on $q$. This dependence comes from needing to ensure that each of the $q$ vectors $\x_k$ are accurately recovered. However, the dependence is pretty weak: when $q < cn^2$, the probability can be further lower bounded by $1-c/n^2$.

%
For the rest of our discussion, assume that $\varepsilon$, $\kappa$ and $\rho$ are fixed. We compare our result with that of TWF initialization \cite{twf}. 
TWF has the best sample complexity, $m \ge c n$, for single vector PR.
Since Theorem \ref{mainthm} provides a guarantee for LRPR-init which exploits the low-rank property of $\X$, when $q/r$ is large, its per column sample complexity is significantly smaller than that of TWF.
For example, if $r = c \log n$ and $q = c r^4 (\log n)^3 = (\log n)^7$, then it needs $\tilde{m} = c\sqrt{n}$ and $m = cn/\log n$ and hence $m_{\total}:=m+\tildm = cn/\log n$. When $q$ is larger, for example, $q = c\sqrt{n}$, it only requires $m_{\total} = c\sqrt{n} r^4 \log^3 n = c\sqrt{n} (\log n)^7$. For $q \ge c n$, just $m_{\total} = c\sqrt{n}$ measurements suffice. This is also backed up by our numerical experiments; see Tables \ref{real_gauss} and \ref{complex_gauss}. Here we used $n=100$ and $r=2$. With as few as $m=5\sqrt{n}$ measurements, when $q=100$, the LRPR-init normalized error is 0.33. When $q=1000$, this error is only 0.1.

When the goal is to only recover $\U$ with subspace error at most $\varepsilon$ (and not the $\x_k$'s), the required lower bounds can be relaxed further. In particular, we have the following corollary.
\begin{corollary} \label{mainthm_cor}
In the setting of Theorem \ref{mainthm}, if $\tilde{m} = 0 $, $m \ge \frac{c\kappa^2 \cdot r^2 \log n}{\varepsilon^2}$ and
$m q \ge   \frac{c \rho^2  \kappa^2 \cdot n r^2  }{\varepsilon^2} $,  then with probability at least $1 - 2\exp(-cn) - \frac{2q}{n^4}$, $\SE(\Uhat,\U) \le  c\varepsilon.$
\end{corollary}
Recall that $\U$ is an $n \times r$ matrix and hence has $nr$ unknowns. From Corollary \ref{mainthm_cor}, for a fixed $\varepsilon$, $\rho$, and $\kappa$, one needs a total of only $mq = c nr^2$ measurements to recover $\U$. When $r$ is small, e.g., $r=c\log n$, this is only slightly more than the minimum required which would be $nr$.

To recover the $\b_k$'s, it follows from Theorem \ref{mainthm} that we need an extra set of $\tilde{m} \ge c\sqrt{n}$ measurements. 
\begin{itemize}
\item The lower bound $\tilde{m} \ge c\sqrt{n}$ can be replaced by $\tilde{m} \ge cn^{1/5}$, or in fact $cn^{1/d}$ for any integer $d \ge 2$, and our result will not change, except for numerical constants.
\item We can even replace $\tilde{m} \ge c\sqrt{n}$ by $\tilde{m} \ge cr \log^4 r$, which is much weaker, but then Theorem \ref{mainthm} will hold with probability lower bounded by only $1 - \frac{8q}{\tilde{m}^8} - 2\exp(-cn) - \frac{8q}{n^4}$.%
\end{itemize}
In Theorem \ref{mainthm}, we also need an extra factor of  $(r \log \tilde{m})^2$ in the lower bounds on $m$ and $mq$  as compared to Corollary \ref{mainthm_cor}. This is needed because our algorithm recovers $\g_k:= \Uhat' \U \b_k$ and sets $\xhat_k = \Uhat \ghat_k$.  Thus, for it to give an accurate enough estimate of $\x_k$, we need to ensure that $\SE(\Uhat,\U)$ is very small so that $\|\Uhat'\U\|$ is close to one. In particular we need $\SE(\Uhat,\U) \le \varepsilon / r \log \tilde{m}$. Guaranteeing this requires a larger lower bound on $mq$ and $m$ than just ensuring $\SE(\Uhat,\U) \le \varepsilon$.

\subsection{Main Results for Random $\X$ - Known rank case}
\label{results_gauss}
First consider an independent zero mean Gaussian model on the $\b_k$'s.%
\begin{sigmodel}\label{gauss_mod}\em
Assume that $\x_k = \U \b_k$ with $\b_k \indepsim \n(0,\Lamk)$, $\Lamk$ diagonal, and $\b_k$'s independent of $\U$.
The matrix $\U$ can follow any probability distribution.
Define
\[
\Lambar: = \frac{1}{q} \sum_k \E[ \b_k \b_k{}'] = \frac{1}{q} \sum_k \Lamk,
\]
let $\lammin$ be its minimum eigenvalue, $\lammax$ its maximum eigenvalue, and $\kappa:= \frac{\lammax}{\lammin}$ its condition number.
Assume also that, for  all $k=1,2,\dots,q$,
\[
\lambda_{k,\max}:= \lambda_{\max}(\Lamk) \le c \lammax.
\]
This is ensured, for example, if $\max_k \lambda_{k,\max} \le c \min_k \lambda_{k,\max}$.
\end{sigmodel}
With this model, notice that $\E[\frac{1}{q} \sum_k \x_k \x_k{}'] \evdeq \U \Lambar \U'$.

In using Model \ref{gauss_mod}, there are two main changes. The first is that we need to apply a law of large numbers result to show that $\frac{1}{q} \sum_k \b_k \b_k{}'$ is close to $\Lambar$ whp. This will hold only when $q$ is large enough, and, hence, our result will also need another lower bound on $q$. The second change is that we need to replace $r \rho \lammax$ by  $r (10 \log n) \lammax$ in the lower bound on $m q$. This is the high probability upper bound on $\|\b_k\|^2 $ under Model \ref{gauss_mod}. Moreover, because of these two changes, the probability of the good event reduces slightly.

\begin{theorem}[Gaussian model] \label{mainthm_gaussian}
In the setting of Theorem \ref{mainthm}, suppose that the $\x_k$'s satisfy Model \ref{gauss_mod}.
For a $\varepsilon < 1$, if
\begin{align*}
& \tilde{m} \ge \frac{ c\sqrt{n}  }{\varepsilon^2}, \ m \ge \frac{c \kappa^2 r^4 (\log \tilde{m})^2}{\varepsilon^2}, \\
& mq \ge \frac{c  \kappa^2  n  r^4 (\log \tilde{m})^2 (\log n)^2 }{\varepsilon^2}, \ q \ge \frac{c \kappa^2 r^3  (\log n) (\log \tilde{m})^2 }{\varepsilon^2},
\end{align*}
then, the conclusions of Theorem \ref{mainthm} hold with probability at least $1- 2\exp(-cn) - \frac{36q}{n^4} - \frac{20}{n^2}$.
\end{theorem}
\begin{proof} See Section \ref{outline}.
\end{proof}
Observe that the lower bound on $q$ in Theorem \ref{mainthm_gaussian} is not very restrictive. From the lower bounds on $\tildm$ and $mq$, $q$ anyway needs to be more than $c r^4 (\log n)^4$ in order to get a lower bound on $m_\total$ that is smaller than $c n$ (which is the best lower bound achievable by a single vector PR method).

As will be evident from the proof of Theorem \ref{mainthm_gaussian}, any random model that ensures that (a) $\max_k \|\b_k\|^2$ is bounded whp, and (b) $\frac{1}{q} \sum_k \b_k \b_k{}'$ is close to $\Lambar$ whp will suffice. For example, even if the $\b_k$'s in Model \ref{gauss_mod} have nonzero and different means, a similar result can be proved. More generally, as we state below, a sub-Gaussian assumption works as well.  The independence assumption on $\b_k$'s may also be weakened to any other assumption that ensures that (b) holds, however we do not pursue it here.


\begin{corollary}[sub-Gaussian model] \label{mainthm_subG}
Let $\x_k = \U \b_k$ with $\b_k$'s being independent of $\U$. Let $\Lambar: = \E[\frac{1}{q} \sum_k \b_k \b_k{}']$, let $\lammax$ be its maximum eigenvalue and $\kappa$ its condition number. Assume that the $\b_k$'s are independent sub-Gaussian random vectors with sub-Gaussian norm bounded by $c\sqrt{\lammax}$.

With this model replacing Model \ref{gauss_mod} on $\X$, Theorem \ref{mainthm_gaussian} holds with probability at least $1- 2\exp(-cn) - \frac{cq}{n^4} - \frac{c}{n^2}.$
\end{corollary}
\begin{proof} In the proof of Theorem \ref{mainthm_gaussian}, only the proofs of Lemmas \ref{D_bnd}, \ref{EY_minus} change.\end{proof} 

\subsection{Main Results - Unknown rank case} \label{unknown_r}
We now turn to the setting where the rank $r$ is unknown and show how Theorem \ref{mainthm} can be modified for this setting. Other results are modified similarly.

Consider the rank estimation approach given in Algorithm \ref{algo_init_disjoint}. 
We have the following corollary.

\begin{corollary}\label{mainthm_unknown_r}
Consider Algorithm \ref{algo_init_disjoint} (unknown $r$ case). Assume the setting of Theorem \ref{mainthm} with $\varepsilon \le 0.001$.
If, in addition, $\kappa \le 10$ and if $\Lambar$ is such that $\lambar_j - \lambar_{j+1} \le 0.9 \lammin$, then,
with the probability given in Theorem \ref{mainthm},
\begin{enumerate}
\item $\hat{r}=r$, and,
\item all conclusions of Theorem \ref{mainthm} hold.
\end{enumerate}
\end{corollary}

Another way to correctly estimate $r$ is via thresholding.
\begin{corollary}\label{mainthm_unknown_r_2}
Consider Algorithm \ref{algo_init_disjoint} with rank estimated as follows. Set $\hat{r}$ as the smallest index $j$ for which $\lambda_j(\Y_U) - \lambda_n(\Y_U) \ge 0.25 \lammin$. 
Assume the setting of Theorem \ref{mainthm} with $\varepsilon \le 0.001$. Then, if $\kappa < 124$, then, with the probability given in Theorem \ref{mainthm},
\begin{enumerate}
\item $\hat{r}=r$, and,
\item all conclusions of Theorem \ref{mainthm} hold.
\end{enumerate}
\end{corollary}

The rank estimation approach of Algorithm \ref{algo_init_disjoint} does not require knowledge of any model parameters.
Hence it is easily applicable for real data (even without training samples being available). However, it works only when consecutive eigenvalues of $\Lambar$ (consecutive nonzero singular values of $\X$) are not too far apart. On the other hand, the thresholding based approach of Corollary \ref{mainthm_unknown_r_2} does not require any extra assumptions beyond those in Theorem \ref{mainthm} and $\kappa < 124$. However it necessitates knowledge of $\lammin$. 



\subsection{Using different measurement vectors for each $\x_k$} \label{indep_meas}
Our problem setting requires that we use a different set of $m$ measurement vectors $\a_{i,k}$ for each column $\x_k$ in order to estimate $\U$. Thus, any application where our algorithms are used needs to apply a total of $m_\total q$ measurement vectors (often, masks), and also needs to store a total of $m_\total  q$ length-$n$ measurement vectors. However, observe that, because we use different measurement vectors and exploit the low-rank property of the matrix $\X$, if $r \le c\log n$, we only need $m_\total q = c n \polylog(n)$ measurement vectors. Here $\polylog(n)$ refers to a polynomial in $(\log n)$.

On the other hand, if we used the same measurement vectors for each column $\x_k$, we would require only $m$ measurement vectors. But we would need $m \ge c n$ such vectors. Since $c n \polylog(n)$ is only a little larger than $c n$, our setting is not much more difficult to implement in practice than the  same measurement vectors' setting.

An advantage of our setting is as follows.
In practice, the region being imaged changes continuously over time. 
Thus, using our approach, one can just acquire a total of $m_\mat$ independent measurements by imaging the region of interest for a certain period of time.
The value of $q$ (and hence of $m=m_\mat/q$) may be decided later depending on the desired tradeoff between temporal resolution and accuracy per pixel\footnote{If the measurements are masked-Fourier, then  this can be done with the constraint that $m$ is an integer multiple of $n$.}. If the changes are gradual, then one can use a smaller value of $q$, but gain in accuracy with a larger $m$.%

\begin{algorithm}
\caption{LRPR+TWF (TWF initialized using LRPR-init)} 
\label{twf_iter_algo}
\begin{enumerate}
\item Initialize $\Xhat^0$ using Algorithm \ref{algo_init}
\item For each $t \ge 0$,  do:
\begin{itemize}
\item
for each $k$, $k=1,2,\dots,q$,  update
\begin{align}
\xhat_k^{t+1} = \xhat_k^t - \frac{\mu}{m} \sum_{i=1}^m \frac{\y_{i,k} - | \a_{i,k}{}' \xhat_k^t|^2}{\a_{i,k}{}'\xhat_k^t} \a_{i,k} \one_{\mathcal{E}_1 \cup \mathcal{E}_2}
\label{truncgd}
\end{align}
where the events $\mathcal{E}_1,\mathcal{E}_2$ are defined in \cite[eq. 28]{twf}.
\end{itemize}

\end{enumerate}
\end{algorithm}

\section{Low Rank PR (LRPR) -  Complete algorithm}
\label{algo_iters}
So far we developed an initialization procedure that directly exploited the low-rank property of $\X$. Here, we explain three possible ways to develop a complete LRPR algorithm. The first, given next, uses LRPR-init to only speed up TWF.

\subsection{LRPR+TWF: speeding up TWF}
Consider the LRPR problem and an application where acquiring measurements is not expensive, but computational power is. In this case, we can use LRPR-init (Algorithm \ref{algo_init}) to jointly initialize all columns of the matrix $\X$, followed by using the best existing vector PR algorithm such as TWF \cite{twf} for recovering each column separately. TWF implements truncated gradient descent for maximizing data likelihood under Poisson measurement noise.
We summarize TWF initialized with LRPR-init (LRPR+TWF) in Algorithm \ref{twf_iter_algo}.
LRPR+TWF still requires $m_\total \ge cn $ measurements per column, but, as we explain next, it needs $c (\log n - 12 \log \log n)$ fewer iterations to converge than the original TWF (basic TWF, Algorithm \ref{twf_iter_algo} initialized using Algorithm \ref{twf_init_algo}).
%
To see this, consider the result of Theorem \ref{mainthm}. Another way to interpret this is as follows. Suppose we are given $m = cn$ and $\tildm = cn$. Then, it is clear that this result holds for any $\varepsilon$ satisfying
\[
\varepsilon^2 \ge c \max \left( \frac{1}{\sqrt{n}},  \frac{\kappa^2 r^4 (\log n)^3}{n}, \frac{\rho^2 \kappa^2  r^4 (\log n)^2 }{q} \right).
\]
If $q \ge c \sqrt{n}$, then this means that $\varepsilon = c \frac{\rho \kappa  r^2 (\log n) }{n^{1/4}}$ works.
Combining this with \cite[Theorem 1]{twf}, we have the following corollary. 

\begin{corollary}[LRPR+TWF] \label{twf_lrpr_cor}
Consider Algorithm \ref{twf_iter_algo}.  If $m = c n$, $\tildm = c n$,  $r \le c \log n$, and
$q \ge c \sqrt{n}$ (but $q \le cn^2$),
then, there exists universal constants $b_1 < 1$ and $c_1$, such that,
 with probability, at least $1 - c/n^2$,
\begin{align*}
\mathrm{NormErr}(\X,\hat\X^t) & \le (1-b_1)^t \mathrm{NormErr}(\X,\hat\X^0) \\
& \le (1-b_1)^t c_1  \frac{\rho \kappa  (\log n)^3}{n^{1/4} }.
\end{align*}
\end{corollary}
From Corollary \ref{twf_lrpr_cor}, to reduce the final error, $\mathrm{NormErr}(\X,\hat\X^T)$, below a given tolerance, $\varepsilon_{\mathrm{fin}}$, Algorithm \ref{twf_iter_algo} needs a total of $T$ iterations, with $T$ satisfying
\[
T \ge  \frac{-\log \varepsilon_{\mathrm{fin}}}{-\log (1-b_1)}  - \frac{0.25  (\log n - 12 \log \log n - \log (c_1  \rho \kappa) ) }{-\log (1-b_1)}.
\]
On the other hand, basic TWF (Algorithm \ref{twf_iter_algo} initialized with TWF initialization, Algorithm \ref{twf_init_algo}) needs
$
T \ge  \frac{-\log \varepsilon_{\mathrm{fin}}}{-\log (1-b_1)}.
$
Thus, using LRPR-init to initialize TWF reduces the number of iterations needed for TWF to converge by $c (\log n - 12 \log \log n).$ However, LRPR-init is also roughly $r$ times more expensive than TWF-init. As seen from Fig.~\ref{TWF_compare}, when $r$ is small, the reduction in number of iterations still results in lower total time taken by LRPR+TWF as compared to basic TWF.%

\subsection{LRPR1: Low Rank PR via projected gradient descent}
The simplest way to develop a complete algorithm that exploits the low rank property of $\X$ is to use a projected gradient descent approach to modify TWF. 
This projects the TWF output at each iteration onto the space of rank $r$ matrices. We summarize the complete LRPR1 approach (projected-TWF initialized with LRPR-init) in Algorithm \ref{twfproj_iter_algo}. When $m$ is small, this results in significantly improved performance over TWF because it exploits the low-rank structure of the matrix $\X$ at each step. For an example, see Fig. \ref{all_compare}.

\begin{algorithm}
\caption{LRPR1: LRPR via projected gradient descent}
\label{twfproj_iter_algo}
\begin{enumerate}
\item Initialize $\Xhat^0$ using Algorithm \ref{algo_init} (LRPR-init).
\item For each $t \ge 0$, do
\begin{enumerate}
\item for each $k$, $k=1,2,\dots,q$, compute $\xhat_k^{t+1}$ using \eqref{truncgd} defined in Algorithm \ref{twf_iter_algo}. Call the resulting matrix $\Xhat^{t+1,TWF}$;
\item project $\Xhat^{t+1,TWF}$ onto the space of rank $r$ matrices to get $\Xhat^{t+1}$.
\end{enumerate}
\end{enumerate}
\end{algorithm}

\subsection{LRPR2: Low Rank PR via Alternating Minimization}
The third and most powerful approach is to modify the entire algorithm to directly exploit the low-rank property of the matrix $\X$, i.e., to use its decomposition as $\X = \U \B$.  This idea can be used to modify TWF or AltMinPhase (Gerchberg-Saxton algorithm) or, in fact, many of the other PR methods from literature, e.g., \cite{bauschke2002phase,bauschke2005new}. As noted by an anonymous reviewer, the last two are significantly faster than Gerchberg-Saxton. TWF is truncated gradient descent to minimize the negative data likelihood under a Poisson noise assumption, where as AltMinPhase is an AltMin approach to minimize the squared loss function (data likelihood under iid Gaussian noise). For noise-free measurements, this distinction is immaterial, and all methods apply.

\input{plane}

Modifying TWF for the set of variables $\U, \B$ needs to be done with care, and needs to include a step that ensures that one of $\|\U\|$ or $\|\B\|$ does not keep increasing. 
An early attempt along these lines is given in \cite{ssp_pr}.

Modifying the AltMin strategy is simpler and we explain it here.  Let $\y_k:=[\y_{1,k}, \y_{2,k}, \dots, \y_{m,k}]'$ and $\A_k:=[\a_{1,k},\a_{2,k},\dots,\a_{m,k}]$. Then $\sqrt{\y_k} = |\A_k{}' \x_k|$. 
Suppose that the phase information were available, i.e., suppose that we had access to a diagonal matrix $\C_k$ so that $\C_k \sqrt{\y_k} = \A_k{}' \x_k$. Then recovering $\X$ from these linear measurements would be an example of a low-rank matrix recovery problem. This itself can be solved by minimizing over $\U$ and $\B$ alternatively as in \cite{lowrank_altmin}. With $\B$ fixed, this is a least squares (LS) recovery problem for $\U$ and vice versa. With estimates of $\U$ and $\B$, we can estimate the phase matrix $\C_k$ as the $\hat\C_k = \mathrm{diag}(\mathrm{phase}(\A_k{}' \Uhat \bhat_k))$.  
The proposed complete algorithm, LRPR2, summarized in Algorithm \ref{LR-AltMin}, alternates between these three steps.
The per iteration cost of the AltMin approach is larger than that of TWFproj iterates and hence LRPR2 is often slower than LRPR1, e.g., see Fig. \ref{all_compare}. However, from numerical experiments, LRPR2 needs the smallest value of $m$ to converge as seen, for example, in Fig. \ref{altmin_compare}. 

\begin{algorithm} 
\caption{\small{LRPR2: LRPR via Alternating Minimization}}\label{LR-AltMin}
\label{amt}
\begin{enumerate}
\item Let $\Uhat$ and $\bhat_k$ denote the output of Algorithm \ref{algo_init}.
\item For $t = 1$ to $T$, repeat the following three steps:
\begin{enumerate}
\item $\Chat_k  \leftarrow  \mathrm{diag}(\mathrm{phase}(\A_k{}'\Uhat  \bhat_k ))$, for $k=1,2,\dots,q$
\item $\Uhat \leftarrow \arg\min_{\tilde\U}  \sum_k \| \Chat_k  \sqrt{\y_k} - \A_k{}' \tilde\U \bhat_k  \|^2$
\item $\bhat_k \leftarrow \arg\min_{\tilde{\b}_k}  \| \Chat_k  \sqrt{\y_k} - \A_k{}' \Uhat  \tilde{\b}_k \|^2$, for $k=1,2,\dots,q$
\end{enumerate}
\item Output $\Uhat$ and $\xhat_k = \Uhat \bhat_k$ for all $k=1,2,\dots, q$.
\end{enumerate}
Steps 2 and 3 involve solving a Least Squares (LS) problem which can be solved in closed form as follows.
\begin{itemize}
\item Step 2: Let $\Uhat_{vec}$ be the columnwise vectorized version of $\Uhat$. Compute
$\Uhat_{vec} = (\sum_k \M_k{}' \M_k)^{-1} \sum_k ( \M_k{}' \Chat_k \sqrt{\y_k} )$ where
$\M_k:=[\A_k{}' (\bhat_k)_1 , \A_k{}' (\bhat_k)_2, \dots, \A_k{}' (\bhat_k)_r]$. Reshape $\Uhat_{vec}$ to get $\Uhat$.  For large sized problems, conjugate gradient for LS (CGLS) is a faster approach to solve the LS problem since it does not require matrix inversion.
\item Step 3: $\bhat_k =(\M'\M)^{-1} \M' \Chat_k \sqrt{\y_k}$ where $\M= \A_k{}' \Uhat$.
\end{itemize}
\end{algorithm}




\input{new_new_expts} 

\section{Proofs of Theorems \ref{mainthm} and \ref{mainthm_gaussian}} \label{outline}

The approach for proving both Theorems \ref{mainthm} and \ref{mainthm_gaussian} is similar.
In Sec.~\ref{key_lem}, we summarize the two results that will be used in our proof - the Davis-Kahan $\sin \theta$ theorem \cite{davis_kahan} and a simple modification of Theorem 5.39 of Vershynin \cite{vershynin}. The $\sin \theta$ theorem  bounds the subspace error between the principal subspaces of a given Hermitian matrix and its perturbed version. The Vershynin result is a probabilistic concentration bound for the empirical covariance matrix of independent sub-Gaussian random vectors. This will be used to bound the terms from the bound obtained by applying the $\sin \theta$ theorem.
In Sec.~\ref{bound_SE}, we bound $\SE(\Uhat,\U)$ both under a deterministic and a random assumption on $\X$.
In Sec.~\ref{bound_xk}, we use this to bound $\dist(\x_k, \xhat_k)$, again under both the deterministic and random settings. 
In Sec.~\ref{proof_mainthm}, we combine these results to prove Theorem \ref{mainthm}. In Sec.~\ref{proof_mainthm_gaussian}, we prove Theorem \ref{mainthm_gaussian}. Finally, we prove the two corollaries for the unknown rank case - Corollaries \ref{mainthm_unknown_r} and \ref{mainthm_unknown_r_2} - in Sec.~\ref{proof_mainthm_unknown_r}.

The derivations in this section use many useful results about sub-Gaussian and sub-exponential r.v.'s and the $\epsilon$-net taken from \cite{vershynin}. These are summarized in Appendix \ref{prelim}. The lemmas that are not proved here are proved in Appendix \ref{proof_lems}.%

\subsection{Davis-Kahan $\sin \theta$ theorem and Vershynin's result} 
\label{key_lem} 

We first state a simple corollary of the Davis-Kahan $\sin \theta$ theorem \cite[Sec. 2]{davis_kahan} that follows from it using Weyl's inequality (see \cite{corpca_nips,rrpcp_perf} for a proof).


\begin{theorem} [$\sin \theta$ theorem \cite{davis_kahan}]  \label{sintheta}
Consider a Hermitian matrix $\cA$ and its perturbed version $\cM$. Define $\cH:=\cM-\cA$.
Let $\bm{E}$ be the matrix of top $r$ eigenvectors of $\cA$, and let $\bm{F}$ be the matrix of top $r$ eigenvectors\footnote{More generally, $\bm{E}$ and $\bm{F}$ can be any matrices whose columns span the space of top $r$ eigenvectors of $\cA$ and $\cM$ respectively.} of $\cM$.
If $\lambda_{r}(\cA) -\lambda_{r+1}(\cA) - \|\cH\| > 0$, then
\beq
\SE(\bF, \bE):= \|(\I-\bF \bF') \bE \| \le \frac{\|\cH\|}{\lambda_{r}(\cA) -\lambda_{r+1}(\cA) - \|\cH\|}.
\nonumber
\eeq
\end{theorem}
In Sec.~\ref{bound_SE}, we will use the above result with $\cM = \Y_U$ and $\cA$ being the expected value of a matrix that is close to it. In Sec.~\ref{bound_xk}, we will use it similarly for $\Y_{b,k}$. 

Theorem \ref{versh} below is a simple generalization of Theorem 5.39 of \cite{vershynin}.
\begin{theorem} \label{versh}
Suppose that $\w_j$, $j=1,2,\dots, N$, are $n$-length independent, sub-Gaussian random vectors with sub-Gaussian norms bounded by $K$. 
\begin{enumerate}
\item For an $\varepsilon < 1$ and a given vector $\z$, with probability (w.p.) $\ge 1 - 2 \exp(-c \varepsilon^2 N)$,
\[
\left| \z' \left(  \frac{1}{N} \sum_j (\w_j \w_j{}' - \E[ \w_j \w_j{}'] )   \right) \z \right|  \le 4\varepsilon K^2 \| \z \|^2.
\]

\item For an $\varepsilon < 1$,   w.p. $\ge 1 - 2 \exp(n \log 9 -c \epsilon^2 N)$,
\[
\left\| \frac{1}{N} \sum_j (\w_j \w_j{}' - \E[ \w_j \w_j{}'] ) \right\| \le  4\varepsilon K^2.
\]
\end{enumerate}
\end{theorem}

\begin{proof} The proof follows that of Theorem 5.39 in \cite{vershynin}. It is given in the Supplementary Material. \end{proof}

\subsubsection{Proving Theorems \ref{mainthm} and \ref{mainthm_gaussian} simultaneously}
Define the following (trivial) model.
\begin{sigmodel} \label{det_mod}
The matrix $\X$ is a deterministic unknown.
\end{sigmodel}
This just makes it simpler to simultaneously obtain subspace error bounds under the assumptions of both Theorems \ref{mainthm} and \ref{mainthm_gaussian}.
Notice that, if we write $\x_k = \U \b_k$, then the definitions of $\Lambar$, $\rho$ and $\kappa$ given in \eqref{def_Lambar} and \eqref{def_rho} in Sec. \ref{results_det} imply that, under Model \ref{det_mod}, $\Lambar = \frac{1}{q} \sum_k \b_k \b_k{}'$, $\kappa$ is its condition number, and $\max_k \|\b_k\|^2=\max_k \|\x_k\|^2 \le r \rho \lammax$.

\subsection{Bounding $\SE(\Uhat,\U)$} \label{bound_SE}

Recall that $\Uhat$ is the matrix of top $r$ eigenvectors of $\Y_U$. To bound $\SE(\Uhat,\U)$ using Theorem \ref{sintheta}, we define a matrix $\bm\Sigma^-$ such that (i) $\U$ is the matrix of its top $r$ eigenvectors; and (ii) there is a significant nonzero gap between its $r$-th and $(r+1)$-th eigenvalues. 
More specifically, we let $\bm\Sigma^-= c_1 \U \Lambar \U' + c_2 \I$ where $c_1$ and $c_2$ are positive constants that are defined later.
 Clearly,  $\lambda_r(\bm\Sigma^-) - \lambda_{r+1}(\bm\Sigma^-)  = c_1 \lammin $.
By Theorem \ref{sintheta}, if $\lambda_r(\bm\Sigma^-) - \lambda_{r+1}(\bm\Sigma^-) > \|\Y_U - \bm\Sigma^-\|$, then,
\begin{equation} \label{sin_theta_bnd}
\SE(\Uhat,\U) 
\le 
\frac{\|\Y_U - \bm\Sigma^-\|}{c_1 \lammin - \|\Y_U - \bm\Sigma^-\|}.
\end{equation}

Thus, all we need now is to specify $\bm\Sigma^-$ and find a high probability upper bound on $\|\Y_U - \bm\Sigma^-\|$. 

To this end, as also done in \cite[Appendix C]{twf}, we first lower and upper bound $\Y_U$ in order to replace $\frac{1}{m} \sum_i \y_{i,k}$ in its indicator function expression by a constant.
Recall that $\Y_U$ is defined in \eqref{def_YU} and that $\frac{1}{m} \sum_i \y_{i,k}  = \x_k{}'(\frac{1}{m} \sum_i \a_{i,k} \a_{i,k}{}') \x_k$. By Fact \ref{fact_versh}, item \ref{K_gauss_vec}, in Appendix \ref{prelim}, $\a_{i,k}$'s are sub-Gaussian with sub-Gaussian norm bounded by $c$.
Thus, using the first part of Theorem \ref{versh}, conditioned on $\x_k$, $|\frac{1}{m} \sum_i \y_{i,k} - \|\x_k\|^2 | \le \epsilon_1 \|\x_k\|^2$ w.p. $ \ge 1 - 2\exp( - c\epsilon_1^2 m)$. The constant multiplying $\epsilon_1$ is moved into the $c$ in the probability. This bound holds for all $k=1,2,\dots,q$ w.p. $ \ge 1 - 2q\exp( - c\epsilon_1^2 m)$.
This implies that, with the same probability, conditioned on $\X$, $\Y^- \preceq  \Y_U \preceq \Y^+ , \ \text{where}$
\begin{align*}
& \Y^- := \frac{1}{m \tot} \sum_{i=1}^m \sum_{k=1}^\tot \w_{i,k}^- \w_{i,k}^- {}', \ \Y^+ := \frac{1}{m \tot} \sum_{i=1}^m \sum_{k=1}^\tot  \w_{i,k}^+ \w_{i,k}^+{}',  \\ 
& \w_{i,k}^- := \left( \a_{i,k}{}'\frac{\x_k}{\|\x_k\|} \right) \a_{i,k} \one_{ \left(\a_{i,k}{}'\frac{\x_k}{\|\x_k\|} \right)^2 \le 9 (1-\epsilon_1)} \ \|\x_k\|, \text{ and}  \\
& \w_{i,k}^+ = \left( \a_{i,k}{}'\frac{\x_k}{\|\x_k\|} \right) \a_{i,k} \one_{ \left(\a_{i,k}{}'\frac{\x_k}{\|\x_k\|} \right)^2 \le 9 (1+\epsilon_1)} \ \|\x_k\|.
\end{align*}
Notice that $\w_{i,k}^+$ is $\w_{i,k}^-$ with $9(1-\epsilon_1)$ replaced by $9(1+\epsilon_1)$ in the indicator function.
The following claim is immediate.
\begin{lemma}\label{one_lem}
Conditioned on $\X$, w.p. $ \ge 1 - 2q\exp( - c\epsilon_1^2 m)$, $\|\Y_U - \Y^-\| \le \|\Y^+ - \Y^-\|$.
\end{lemma}
Define
\[
\bm\Sigma^- :=\E[\Y^-] \text{ and } \bm\Sigma^+ :=\E[\Y^+].
\]
%
We obtain expressions for these in the next lemma.
\begin{lemma} \label{expec_Y}
Let $\xi \sim \n(0,1)$. Define
\begin{align*}
& \beta_1^- = \beta_1^-(\epsilon_1) :=  \E[ (\xi^4 - \xi^2) \one_{\xi^2 \le 9(1-\epsilon_1)}], \\
& \beta_2^- = \beta_2^-(\epsilon_1):= \E[ \xi^2 \one_{\xi^2 \le 9(1-\epsilon_1)}].
\end{align*}
Under both Models \ref{det_mod} and \ref{gauss_mod},
\begin{align*}
& \E[\Y^- | \X]=\beta_1^- \U \left( \frac{1}{q} \sum_k \b_k \b_k{}' \right) \U' + \beta_2^- \left( \frac{1}{q} \sum_k \|\b_k\|^2 \right) \I, \\
& \text{and } \bm\Sigma^- =  \beta_1^- \U \Lambar \U' + \beta_2^- \mathrm{trace}(\Lambar) \I.
\end{align*}
The matrices $\E[\Y^+ | \X]$ and $\bm\Sigma^+$ have similar expressions where we replace $\beta_i^-$ by $\beta_i^+$, $i=1,2$. For defining $\beta_i^+$, replace $9(1-\epsilon_1)$ in the indicator function by  $9(1+\epsilon_1)$.
\end{lemma}

By the triangle inequality and Lemma \ref{one_lem}, conditioned on $\X$, w.p. $ \ge 1 - 2q\exp( - c\epsilon_1^2 m)$,
\begin{align*}
\|\Y_U - \bm\Sigma^-\| & \le \|\Y_U - \Y^-\| + \|\Y^- - \bm\Sigma^-\|  \\
& \le \|\Y^+ - \Y^-\| + \|\Y^- - \bm\Sigma^-\| \\
& \le 2\|\Y^- - \bm\Sigma^-\| + \|\Y^+ - \bm\Sigma^+\|  + \|\bm\Sigma^+ - \bm\Sigma^-\|.
\end{align*}
To bound $\|\Y^- - \bm\Sigma^-\|$, we first bound $\|\Y^- - \E[\Y^-| \X]\|$ using the second claim of Theorem \ref{versh}. We bound $\|\Y^+ - \bm\Sigma^+\|$ similarly.
\begin{lemma}
\label{Y_minus}
Conditioned on $\X$, w.p. $\ge 1 - 2\exp(n \log 9 - \epsilon_2^2 m q )$,
\[
\| \Y^- - \E[\Y^- | \X]\| \ \le \epsilon_2 \max_{k} \|\b_k\|^2.
\]
The same bound holds with the same probability for $\| \Y^+ - \E[\Y^+ | \X]\|$.
\end{lemma}

\begin{remark} \em
Since the claim of Lemma \ref{Y_minus} holds with the same probability lower bound for all $\X$, it also holds with the same probability lower bound if we average over $\X$. The same is true for Lemma \ref{one_lem}.
\end{remark}

The next lemma bounds $ \max_{k} \|\b_k\|^2 =  \max_{k} \|\x_k\|^2$.

\begin{lemma}\label{D_bnd}
Under Model \ref{det_mod}, $ \max_{k} \|\b_k\|^2 \le r \rho \lammax$.
Under Model \ref{gauss_mod}, w.p. $\ge 1 -  2q / n^4$, $ \max_{k} \|\b_k\|^2 \le r (10 \log n) \lammax$.
\end{lemma}

Under Model \ref{det_mod}, $\bm\Sigma^- = \E[\Y^- | \X]$ and so $\|\E[\Y^- | \X] - \bm\Sigma^-\| = 0$.
Under Model \ref{gauss_mod}, we use the second claim of Theorem \ref{versh}, to bound $\|\E[\Y^- | \X] - \bm\Sigma^-\|$ as follows.
\begin{lemma}\label{EY_minus}
Under Model \ref{det_mod}, $\|\E[\Y^- | \X] - \bm\Sigma^-\| = 0$.
Under Model \ref{gauss_mod}, w.p. $\ge 1 - 2\exp(r \log 9 - c\epsilon_3^2 q) - 18 \exp(- c\epsilon_3^2 \frac{q}{r})$,
\[
\|\E[\Y^- | \X] - \bm\Sigma^-\| \le  \epsilon_3  \lammax.
\]
\end{lemma}

 Combining Lemmas \ref{Y_minus}, \ref{D_bnd} and \ref{EY_minus}, we can bound $\|\Y^- - \bm\Sigma^-\|$ under both models. We get the same bound on $\|\Y^+ - \bm\Sigma^+\|$ as well.

Finally, we bound $\|\bm\Sigma^+ - \bm\Sigma^-\|$ using the fact that, for all $\xi > \xi_0$, $\xi^d e^{-\frac{\xi^2}{4}} \le b$ for any $d > 1$.
\begin{lemma}
\label{Sigma_minus}
If $\epsilon_1 \le 1/9$, then
\begin{align*}
\|\bm\Sigma^+ - \bm\Sigma^-\|
& \le [(\beta_1^+(\epsilon_1) - \beta_1^-(\epsilon_1)) + (\beta_2^+(\epsilon_1) - \beta_2^-(\epsilon_1)) r] \lammax  \\
& \le  30 r \epsilon_1  \lammax
\end{align*}
\end{lemma}

Combining the above bounds, we conclude the following.
\begin{corollary} \label{final_YU_bnd}
Let
\begin{align}
& p_{U,1}: =  2q\exp( - c\epsilon_1^2 m) +  4\exp(n \log 9 - c\epsilon_2^2 m q ), \nonumber \\
& p_{U,2}: =  
 p_{U,1} + \frac{4q}{n^4} + 4\exp(r \log 9 - c\epsilon_3^2 q) + 36 \exp(- c\epsilon_3^2 \frac{q}{r}),  \nonumber \\
& \epsilon_{U,1}: = 3  r \epsilon_2 \rho + 30 r \epsilon_1, \nonumber \\
& \epsilon_{U,2}: = 3  r \epsilon_2 (10 \log n) + 30 r \epsilon_1 + 3  \epsilon_3.
\label{def_p_eps}
\end{align}
Then,
\begin{enumerate}
\item under Model \ref{det_mod}, w.p. $ \ge 1 - p_{U,1}$, $\|{\Y}_U - \bm\Sigma^-\| \le  \epsilon_{U,1} \lammax$; and

\item under Model \ref{gauss_mod}, w.p. $ \ge 1 - p_{U,2}$, $\|{\Y}_U - \bm\Sigma^-\| \le  \epsilon_{U,2} \lammax.$
\end{enumerate}
\end{corollary}
Finally, to bound the subspace error of $\Uhat$, we also require a lower bound on $\beta_1^-$. This follows easily using the fact that, for all $\xi > \xi_0$, $\xi^d e^{-\frac{\xi^2}{4}} \le b$ for any $d > 1$.
\begin{lemma}\label{beta1_minus_bnd}
If $\epsilon_1 \le 1/9$, then
\[
\beta_1^-(\epsilon_1) = 2 - \E[(\xi^4 - \xi^2) \one_{\xi^2 \ge (9-9\epsilon_1)}]  \ge 0.5.
\]
\end{lemma}
Applying the $\sin\theta$ theorem, Theorem \ref{sintheta}, and the last two claims above, we get the following result.
\begin{corollary}[$\SE(\Uhat,\U)$ bound] \label{SE_bnd_fin}
Let Model 1 be Model \ref{det_mod} and Model 2 be Model \ref{gauss_mod}.
If $\kappa \epsilon_{U,d} < 1/16$ then, under Model $d$,
w.p. $ \ge 1 -  p_{U,d}$, 
\[
\delta_U:= \SE(\Uhat,\U) \le  2.3  \kappa \epsilon_{U,d}.
\]
\end{corollary}

\begin{proof}
Using Theorem \ref{sintheta}, 
\[
\SE(\Uhat,\U) \le
\frac{\|\Y_U - \bm\Sigma^-\|}{\beta_1^- \lammin - \|\Y_U - \bm\Sigma^-\|}.
\]
Since $\epsilon_1 \le \epsilon_{U,d} \le \kappa \epsilon_{U,d} \le 1/16$, using Lemma \ref{beta1_minus_bnd}, $\beta_1^- \ge 0.5$. Thus,
 under Model $d$, w.p. $ \ge 1 - p_{U,d}$,
\[
\delta_U:= \SE(\Uhat,\U) \le  \frac{  \frac{\kappa\epsilon_{U,d}}{0.5} }{ 1 - \frac{\kappa \epsilon_{U,d}}{0.5} } <  \frac{8}{7}  \frac{1}{0.5}   \kappa \epsilon_{U,d} < 2.3  \kappa \epsilon_{U,d},
\]
completing the proof.
\end{proof}

\subsection{Bounding $\dist(\x_k,\xhat_k)$}\label{bound_xk}

Recall that $\delta_U:=\SE(\Uhat,\U)$ was bounded above. Here, we bound $\dist(\x_k,\xhat_k)$'s in terms of $\delta_U$ and other quantities.

\begin{remark} \em
For notational simplicity, we let $\x = \x_k$, $\b:= \b_k$, $\y_i:=\y_{i,k}^{new}$, $\a_i:=\a_{i}^{new}$, ${\Y}_{\b}:= {\Y}_{\b,k}$ defined in \eqref{def_Yb} in Algorithm \ref{algo_init_disjoint}. Since the different $\b_k$'s are recovered separately, but using the same technique, this notation does not cause any confusion at most places. Where it does, we clarify. 

In this section, we state all results conditioned on $\x$ and $\Uhat$. Under this conditioning, in all our claims, the probability of the desired event is lower bounded by a value that does not depend on $\x$ or $\Uhat$. Thus, the same probability lower bound holds even when we average over $\x$ and $\Uhat$ (holds unconditionally).
\end{remark}


Notice that $\x = \U \b$ can be rewritten as
\begin{align*}
& \x = \Uhat \g + \e, \ \text{where} \\
& \g:= \Uhat' \x = \Uhat'\U \b,  \ \e:=  (\I - \Uhat \Uhat{}') \x. 
\end{align*}
We can further split $\g$ as $\g = \gd \ng $ where $\ng = \|\g\|$ and $\gd = \g/\ng$.
Recall that we estimate $\x$ as 
\[
\xhat = \Uhat \ghat =\Uhat \ghatd \hatng,
\]
where $\ghatd$ is the top eigenvector of $\Y_b$ and $\hatng = \sqrt{ \frac{1}{\tildm} \sum_i \y_i }$.
The following fact is immediate.
\begin{fact} \em \label{ng_bnd}
The vector $\e$ and the scalar $\ng:=\|\g\|$ satisfy
$\|\e\| \le \delta_U \|\b\|$, and\footnote{using $\ng = \|\Uhat \g \|  \ge \|\x\| - \|\e\| \ge (1- \delta_U) \|\b\|$,}
$(1- \delta_U) \|\b\| \le \ng \le \|\b\|$.
\end{fact}


Using the definition of $\dist$, it is easy to see that\footnote{$\dist(\x_1, \x_2) = \min_{\phi \in [0,2\pi]} \|\x_1 - e^{j \phi} (\x_2 + \x_3 - \x_3)\| \le \min_{\phi \in [0,2\pi]} (\|\x_1 - e^{j \phi}\x_3\| + \|\x_2 - \x_3\|) = \dist(\x_1, \x_3) + \|\x_2 - \x_3\|$.}
\beq
\dist(\x_1, \x_2) \le \dist(\x_1, \x_3) + \|\x_3 - \x_2\|.
\label{dist_fact}
\eeq
From \eqref{dist_fact} and Fact \ref{ng_bnd},
\begin{itemize}
\item $\dist(\x,\xhat) \le  \delta_U \| \b\| + \dist(\g,\ghat)$;
\item $\dist(\g,\ghat) \le 
\|\b\|\dist(\gd,  \ghatd) + |\ng - \hatng |$;
\item $\dist(\gd,  \ghatd)^2 = 2(1 - |\ghatd' \gd |)$, and $\SE(\ghatd,\gd) = \|\gd -  \ghatd \ghatd' \gd \| \ge 1 - |\ghatd' \gd |$. Thus,
\end{itemize}
\beq\label{dist_bnd}
\dist(\x,\xhat) 
\le \delta_U \| \b\|  + \sqrt{2 \SE(\ghatd,\gd)} \|\b\|  + |\ng - \hatng|
\eeq
We now need to bound $\SE(\ghatd,\gd)$ and $|\ng - \hatng|$. Using the first claim of Theorem \ref{versh}, we can bound the latter as follows.
\begin{lemma}[$| \hatng - \ng |$ bound] \label{ng_bnd_lem}
Conditioned on $\x$ and $\Uhat$, w.p. $\ge 1- 2\exp(-c\epsilon_4^2 \tilde{m})$, $| \hatng - \ng | \le \epsilon_4 \|\b\| + \delta_U \| \b \|$.

\noindent This holds for all $\nu_k$'s, $k=1,2,\dots,q$, w.p. $\ge 1- 2q \exp(-c\epsilon_4^2 \tilde{m})$.
\end{lemma}

To bound $\SE(\ghatd,\gd)$, we use Theorem \ref{sintheta} with $\cM = \Y_b$. To define a matrix $\cA$, whose top eigenvector is $\gd$, 
let
\begin{align}
& \ta_i: = \Uhat'\a_i, \\ 
& \Y_g:=\frac{1}{\tilde{m}} \sum_{i=1}^{\tilde{m}}  (\ta_i{}' \g)^2 \ta_i \ta_i{}'.
\label{def_Yg}
\end{align}
It is easy to see that, conditioned on $\Uhat$, $\ta_i \iidsim \n(0,\I)$.
By letting $\g$ be the first column of $\I$ and using rotation invariance of $\ta_i$ \cite[Lemma A.1]{wf},
\[
\E[\Y_g | \x, \Uhat] = 2\g \g' + \|\g\|^2 \I.
\]
Clearly, the top eigenvector of this matrix is proportional to $\g$ and the desired eigen-gap is $2\|\g\|^2 \ge 2(1-\delta_U)^2 \|\b\|^2$.
So, we can use $\cA =  2\g \g' + \|\g\|^2 \I$ for applying Theorem \ref{sintheta}. It remains to bound $\|\Y_b - (2\g \g' + \|\g\|^2 \I)\|$.

Recall that $\Y_b = \frac{1}{\tilde{m}} \sum_{i=1}^{\tilde{m}}  (\a_i{}'\x)^2  \ta_i \ta_i{}'$. Thus,
\begin{align} \label{Yb_bnd}
& \|\Y_b -  ( 2\g \g' + \|\g\|^2 \I )  \|  \nonumber \\
&  \le \|\Y_g -  ( 2\g \g' + \|\g\|^2 \I ) \| + \|\Y_{e,1}\| + 2\|\Y_{e,2}\|,
\end{align}
\text{where}
\begin{align*}
& \Y_{e,1}:=\frac{1}{\tildm} \sum_{i=1}^\tildm  \e' \a_i \a_i{}' \e  \ta_i \ta_i{}', \ \text{and} \nonumber \\
& \Y_{e,2}:= \frac{1}{\tildm} \sum_{i=1}^\tildm  \e' \a_i \ta_i{}' \g   \ta_i \ta_i{}'.
\end{align*}
%
%
To bound $\|\Y_g -  ( 2\g \g' + \|\g\|^2 \I ) \|$, we use following modification of Theorem 4.1 of \cite{pr_altmin}.
\begin{lemma} \label{Yg_bnd} [modified version of Theorem 4.1 of \cite{pr_altmin}]
Let $\Y_g$ be as defined in \eqref{def_Yg}. Recall that $\g:=\Uhat'\U \b$ is an $r$-length vector and, conditioned on $\Uhat$, $\ta_i \iidsim \n(0,\I)$.
If $\tilde{m} > c r \log^4 r / \epsilon_5^2$, then, conditioned on $\x$ and $\Uhat$, w.p. $\ge 1  - 4 / \tilde{m}^8$,
\begin{align*}
& \|\Y_{\g} - ( 2\g \g' + \|\g\|^2 \I ) \|   \\
& \le \left( \sqrt{ \frac{8000  r \log^4 \tilde{m}}{\tilde{m}} } + \frac{\sqrt{2}}{\tilde{m}^4} \right) \|\g\|^2 \le 2{\epsilon_5} \|\g\|^2.
\end{align*}
This holds for all $\g_k$'s, for $k=1,2,\dots,q$, w.p. $\ge 1  - 4q / \tilde{m}^8$.
\end{lemma}

Next, consider $\|\Y_{e,1}\|$. In the argument below, everything is conditioned on $\Uhat$ and $\x$.
Suppose that $\tildm > r$.
Using Cauchy-Schwartz for matrices, Theorem \ref{CSmat}, with $\bm\tilde{X}_i = \e' \a_i \a_i{}' \e$ and $\bm\tilde{Y}_i = \ta_i \ta_i{}' $, and simplifying the resulting bounds, we get
\begin{align*}
\|\Y_{e,1}\|^2 &  \\
\le &  \left\| \frac{1}{\tilde{m}} \sum_i \e' \a_i \a_i{}' \e \e' \a_i \a_i' \e \right\| \cdot \left\|\frac{1}{\tilde{m}} \sum_i  \ta_i \ta_i{}' \ta_i \ta_i{}' \right\|  \\
\le &  \max_i \a_i{}' \e \e' \a_i  \  \left\| \frac{1}{\tilde{m}} \sum_i \e' \a_i  \a_i{}' \e \right\|  \  \times    \\
    & \max_i \ta_i{}' \ta_i \   \left\|\frac{1}{\tildm} \sum_i  \ta_i  \ta_i{}' \right\|  \\
\le   &  (80 r \log \tilde{m}) \|\b\|^2  (1+\epsilon_4) \|\e\|^2  (20 r \log \tildm )  (1+\epsilon_4) \\
 \le  & 1600 (r \log \tilde{m})^2  \delta_U^2 (1+\epsilon_4)^2 \|\b\|^4,
\end{align*}
w.p. $\ge 1 -  4/\tilde{m}^8  - 2\exp(-c\epsilon_4^2 \tilde{m}) - 2\exp(r \log 9 - c\epsilon_4^2 \tilde{m})$.

The first inequality is by Cauchy-Schwartz (Theorem \ref{CSmat}). The second one pulls out the scalar $ \max_i \a_i{}' \e \e' \a_i$ from the first summation and the scalar $ \max_i \ta_i{}' \ta_i$ from the second summation.
The third inequality relied on the following arguments to bound the four terms.  It used (i) Theorem \ref{versh}, part 1 with $\z = \e$ to bound $\left\| \e' (\frac{1}{\tilde{m}} \sum_i  \a_i  \a_i' ) \e \right\|$; (ii) Theorem \ref{versh}, part 2 for bounding $\frac{1}{\tilde{m}} \sum_i \ta_i  \ta_i{}' $; and (iii) Fact \ref{fact_versh}, item \ref{rayleigh_type_bnd} in Appendix \ref{prelim} to bound $\max_i \|\ta_i\|^2$ by $20 r \log \tilde{m}$ w.p. $\ge 1- 2/\tilde{m}^8$ (since $\tilde{m} > r$). This was possible because conditioned on $\Uhat$, $\ta_i \sim \n(0,\I)$. (iv) Finally, it used the triangle inequality, Fact \ref{fact_versh}, item \ref{rayleigh_type_bnd} and the fact that $\ta_i = \Uhat' \a_i \sim \n(0,\I)$ and $\U'\a_i \sim \n(0,\I)$ to bound
\begin{align*}
\max_{i}  |\a_i{}' \e| & \le \max_i \|\a_i{}'\U\| \|\b\| + \max_i \| \a_i{}'\Uhat\| \|\Uhat' \U\| \|\b\| \\
& \le 2 \sqrt{20 r \log \tilde{m}} \|\b\|,
\end{align*}
w.p. $\ge 1- 2/\tilde{m}^8$.
The fourth inequality in the bound on $\|\Y_{e,1}\|^2$ follows using $\|\e\| \le \delta_U \|\b\|$.

The value of $\|\Y_{e,2}\|$ can be bounded in a similar fashion:%
\begin{align*}
\|\Y_{e,2}\|^2 &  \le \| \frac{1}{\tildm} \sum_i \e' \a_i \ta_i{}' \g \g' \ta_i \a_i' \e \|  \ \|\frac{1}{\tildm} \sum_i  \ta_i \ta_i{}' \ta_i \ta_i{}'\| \\
& \le 400 (r \log \tildm)^2 \delta_U^2 (1 + \epsilon_4)^2 \|\b\|^4,
\end{align*}
with the same probability. The main difference here is that we  bound $\max_i (\ta_i{}' \g)^2$ instead of $\max_i (\a_i{}' \e)^2$. To do this, we use $\max_i (\ta_i{}' \g)^2 \le \max_i \|\ta_i\|^2 \|\g\|^2 \le 20 r (\log \tilde{m}) \|\b\|^2$ w.p. $\ge 1- 2/\tilde{m}^8$.
Thus, we have the following lemma.
\begin{lemma}\label{Ye_bnd}
Suppose that $\tilde{m} > r$.
Conditioned on $\x$ and $\Uhat$, w.p. $\ge 1 -  6/\tilde{m}^8  - 4\exp(-c\epsilon_4^2 \tilde{m}) - 2\exp(r \log 9 - c\epsilon_4^2 \tilde{m})$, 
\beqa
&& \|\Y_{e,1}\| \le  40 (r \log \tilde{m})  \delta_U (1+\epsilon_4) \|\b\|^2,  \nonumber \\ 
&& \|\Y_{e,2}\| \le 20 (r \log \tilde{m})  \delta_U (1+\epsilon_4) \|\b\|^2.
\eeqa
This holds for all $\g_k$'s for $k=1,2,\dots,q$, w.p. $\ge 1 -  6q/\tilde{m}^8  - 4q \exp(-c\epsilon_4^2 \tilde{m}) - 2 \exp(r \log 9 - c\epsilon_4^2 \tilde{m}).$
\end{lemma}

Let
\beq
p_g:= \frac{10 q}{\tilde{m}^8}  + 4q \exp(-c\epsilon_4^2 \tilde{m}) + 2\exp(r \log 9 - c\epsilon_4^2 \tilde{m}).
\label{def_p_g}
\eeq
Using \eqref{Yb_bnd} and the bounds from Lemmas \ref{Yg_bnd} and \ref{Ye_bnd}, we conclude the following.
If $\tilde{m} \ge \frac{1}{\epsilon_5^2}c r \log^4 r$, then, conditioned on $\X$ and $\Uhat$, w.p. $\ge 1- p_g$, for all $k=1,2,\dots,q$,
\[
\|\Y_{b,k} - (2\g_k \g_k{}' + \|\g_k\|^2 \I)  \| \le (2\epsilon_5 + 80 r \log \tilde{m} (1 + \epsilon_4) \delta_U ) \|\b_k\|^2.
\]
Using Theorem \ref{sintheta} and Fact \ref{ng_bnd}, with the same probability,
\begin{eqnarray}
\SE(\ghatd_k,\gd_k)
& \le & \frac{\|\Y_{b,k} - (2\g_k  \g_k{}' + \|\g_k \|^2 \I)  \|}{2\|\g_k \|^2 - \|\Y_{b,k} - (2\g_k  \g_k{}' + \|\g_k \|^2 \I)  \|} \nonumber \\
& \le &
 \frac{(2\epsilon_5 + 80 r \log \tilde{m} (1 + \epsilon_4) \delta_U )}{(2-2\delta_U - (2\epsilon_5 + 80 r \log \tilde{m} (1 + \epsilon_4) \delta_U )}. \nonumber
\end{eqnarray}
If the numerator is smaller than  $1-2\delta_U$, then 
\[
\SE(\ghatd_k,\gd_k) \le  (2\epsilon_5 +  80 r \log \tilde{m} (1 + \epsilon_4) \delta_U ).
\]
As before, we can average over $\X$ and $\Uhat$ and still get all the events above to hold with the same probability.
Using the above bound, \eqref{dist_bnd}, and Lemma \ref{ng_bnd_lem}, we get the following.
\begin{corollary}[$\dist(\xhat_k,\x_k)$ bound]\label{SE_gd_bnd} \label{fin_dist_bnd}
If $\tilde{m} \ge \frac{1}{\epsilon_5^2}c r \log^4 r$, w.p. $\ge 1- p_g - 2q\exp(-c\epsilon_4^2 \tilde{m})$, $\dist(\xhat_k,\x_k)$ is upper bounded by
\[
(\epsilon_4 +2\delta_U + \sqrt{ 2 (2\epsilon_5 + 80 r \log \tilde{m} (1 + \epsilon_4) \delta_U ) } ) \|\b_k\|,
\]
for all $k=1,2,\dots,q$, if $2\epsilon_5 +  r \log \tilde{m} (1 + \epsilon_4) \delta_U  < 1 - 2 \delta_U$.
\end{corollary}

\subsection{Proof of Theorem \ref{mainthm}} \label{proof_mainthm}

%
%
Combining Corollaries \ref{SE_bnd_fin} and  \ref{SE_gd_bnd}, we can conclude the following.
If $\tilde{m} \ge c r \log^4 r/\epsilon_5^2$, then, w.p. $\ge 1- p_{U,1} - p_g - 2q \exp(- c\epsilon_4^2 \tilde{m})$, $\dist(\xhat_k,\x_k)$ is bounded by
\[
\left(\epsilon_4 + 4.6 \kappa \epsilon_{U,1} + \sqrt{ 2 (2\epsilon_5 + 184 r \log \tilde{m} (1 + \epsilon_4) \kappa  \epsilon_{U,1} ) } \right) \|\b_k\|
\]
for all $k=1,2,\dots,q$, as long as 
$\kappa  \epsilon_{U,1}  \le 1/16$ (this automatically implies $\epsilon_1<1/9$) and $(2\epsilon_5 +  184 r \log \tilde{m} (1 + \epsilon_4) \kappa  \epsilon_{U,1}) \le 1 - 4.6 \kappa  \epsilon_{U,1}$.
%
%

Recall that $\epsilon_{U,1} = r \rho \epsilon_2  + r \epsilon_1$.
Thus to get $\dist(\xhat_k,\x_k)$ below $c\sqrt{\varepsilon}\|\b_k\|$, for an $\varepsilon<1$, we set  
\[
\epsilon_1 = \frac{\varepsilon}{150\kappa r^2 \log \tilde{m}}, \
\epsilon_2 = \frac{\varepsilon}{15\kappa \rho r^2 \log \tilde{m}}, \
\epsilon_4 = \sqrt{\varepsilon}, \
\epsilon_5 = \varepsilon/5.
\]
With these choices notice that, if $\tilde{m} \ge 3$ and $r \ge 3$, then 
$\kappa  \epsilon_{U,1} = \kappa( 3r \rho \epsilon_2 + 30 r \epsilon_1 ) =  2\varepsilon /(5r\log \tilde{m})
< 1/16$; and $2\epsilon_5 +  r \log \tilde{m} (1 + \epsilon_4) \kappa  \epsilon_{U,1} \le  1 - 2/16$.

Using the expressions for $p_{U,1}$ and $p_g$, to get the probability of the desired event below $1 - 2\exp(-cn) - 32q/n^4$, we need
\begin{align*}
& \frac{\varepsilon^2}{c\kappa^2 r^4 \log^2 \tilde{m}} m \ge 4 \log n,   \
\frac{\varepsilon^2}{c\kappa^2 \rho^2 r^4 \log^2 \tilde{m}} m q \ge c n,  \\
& \tilde{m} \ge c\sqrt{n}, \ \tilde{m} \ge \frac{c r \log^4 r }{ \varepsilon^2}, \\
& \varepsilon \tilde{m} \ge 4 \log n, \    \tilde{m} \ge \frac{(4 \log n + (\log 9) r)}{\varepsilon}.
\end{align*}
We obtained these bounds by taking each probability term and finding a lower bound on $\tilde{m}$, $m$ or $mq$ to get it below either $cq/n^4$ or $2\exp(-cn)$.
Theorem \ref{mainthm} assumes $r \le n^{1/5}$. Thus, $r \log^4 r \le c\sqrt{n}$, $r \le c \sqrt{n}$. For large $n$, $\log n \le c\sqrt{n}$.

\subsection{Proof of Theorem \ref{mainthm_gaussian}} \label{proof_mainthm_gaussian}
We use the same approach as above. With Model \ref{gauss_mod}, both $\epsilon_U$ and $p_U$ are larger.
We have $\epsilon_U = \epsilon_{U,2}$ which is equal to $3 \epsilon_3$ plus $\epsilon_{U,1}$ with $\rho$ replaced by $10\log n$. Also, $p_U = p_{U,2} = p_{U,1} +2q/n^4 + 2\exp(r\log 9 - c\epsilon_3^2 q) + 18 \exp(- c\epsilon_3^2 \frac{q}{r})$.

Thus, two things change in the conditions required to get the error below $c\sqrt{\varepsilon}$ w.p. $\ge 1 - 2\exp(-cn) - 16q/n^4 - 2q/n^4 - 20/n^4 $.
First, $\rho^2$ is replaced by $(10\log n)^2$ in the lower bound on $mq$. Second, we set
$\epsilon_3  = \frac{\varepsilon}{\kappa r \log \tilde{m}}$. With this, to get $2\exp(r\log 9 - c\epsilon_3^2 q) < 2/n^4$ and $18 \exp(- c\epsilon_3^2 \frac{q}{r}) < 18/n^4$, we need the assumed lower bound on $q$.


\subsection{Proof of Corollary \ref{mainthm_unknown_r} and Corollary \ref{mainthm_unknown_r_2}} \label{proof_mainthm_unknown_r}
From the proof given in the previous subsections, 
we need $\hat{r}=r$ in order to apply the $\sin \theta$ theorem (Theorem \ref{sintheta}) to bound $\SE(\Uhat,\U)$ in Corollary \ref{SE_bnd_fin}.

Using Corollary \ref{final_YU_bnd}, w.p. $\ge 1 - p_{U,1}$,
\begin{align}
& \|\Y_U - \bm\Sigma^-\| \le \epsilon_{U,1} \lammax = \kappa \epsilon_{U,1} \lammin, \ \text{where} \nonumber \\
& \bm\Sigma^- := \beta_1^-(\epsilon_1) \U \Lambar \U' + \beta_2^-(\epsilon_1) \trace(\Lambar) \I,
\label{bnd_YU_Sig}
\end{align}
and $\beta_1^-(\epsilon_1), \beta_2^-(\epsilon_1)$ are defined in Lemma \ref{expec_Y}. Let $\beta_1^- :=\beta_1^-(\epsilon_1)$.
%
%
Suppose that $\varepsilon \le 0.001$. From Sec. \ref{proof_mainthm}, $\epsilon_1 \le \epsilon_{U,1} \le \varepsilon \le 0.001$. Using $\kappa \le 10$, $2 \kappa \epsilon_{U,1} \le 0.02$.

Thus, from \eqref{bnd_YU_Sig}, Weyl's inequality \cite{hornjohnson}, and $\lambar_j - \lambar_{j+1} \le 0.9 \lammin$, we conclude the following: for a $j < r$ and a $j' > r$, w.p. $\ge 1 - p_{U,1}$,
\begin{align*}
 \lambda_r(\Y_U) - \lambda_{r+1}(\Y_U) & \ge (\beta_1^- - 2 \kappa \epsilon_{U,1} ) \lammin \\
                                      & \ge (\beta_1^- - 0.02 ) \lammin,  \\
 \lambda_j(\Y_U) - \lambda_{j+1}(\Y_U)&
 \le \lambda_j(\bm\Sigma^-) - \lambda_{j+1}(\bm\Sigma^-) + 2  \kappa \epsilon_{U,1} \lammin   \\
& = \beta_1^- (\lambar_j -  \lambar_{j+1}) +  2  \kappa \epsilon_{U,1} \lammin, \\
& \le (0.9 \beta_1^-  +  0.02) \lammin, \text{ and} \\
 \lambda_{j'}(\Y_U) - \lambda_{j'+1}(\Y_U) & \le \lambda_{j'}(\bm\Sigma^-) -  \lambda_{j'+1}(\bm\Sigma^-)  +  2  \kappa \epsilon_{U,1} \lammin \\
 & = 0 + 2 \kappa \epsilon_{U,1} \lammin \le 0.02 \lammin.
\end{align*}

By Lemma \ref{beta1_minus_bnd}, $\beta_1^- \ge 0.5$. Using this, we conclude that
\begin{align*}
\lambda_r(\Y_U) - \lambda_{r+1}(\Y_U) & \ge (\beta_1^-  - 0.02) \lammin \\
 & > (0.9  \beta_1^- + 0.02) \lammin \\
 & \ge \lambda_j(\Y_U) - \lambda_{j+1}(\Y_U), \text{ and} \\
\lambda_r(\Y_U) - \lambda_{r+1}(\Y_U)
&  \ge (\beta_1^-  - 0.02) \lammin   \\ 
&  >  0.02 \lammin \ge  \lambda_{j'}(\Y_U) - \lambda_{j'+1}(\Y_U)
\end{align*}
for a $j<r$ and a $j' >r$. The second row used $0.1 \beta_1^-  > 0.04$ and the last row used $\beta_1^-  - 0.02 > 0.02$.

Thus, if $\kappa \le 10$, and $\lambar_j - \lambar_{j+1} \le 0.9 \lammin$, then, under the assumptions of Theorem \ref{mainthm},
w.p. $\ge 1 - p_{U,1}$, $\lambda_j(\Y_U) - \lambda_{j+1}(\Y_U)$ is largest for $j=r$, i.e., $\hat{r}=r$.
Using this and then proceeding exactly as before, we obtain Corollary \ref{mainthm_unknown_r}. 

To get Corollary \ref{mainthm_unknown_r_2}, use \eqref{bnd_YU_Sig}, Weyl's inequality \cite{hornjohnson}, and $\kappa \le 124$, to argue that, w.p. $\ge 1 - p_{U,1}$,  for any $j > r$,
\begin{align*}
& \lambda_r(\Y_U) - \lambda_{n}(\Y_U) \ge (\beta_1^-  - 0.002 \kappa) \lammin \ge (0.5  - 0.246)\lammin, \\ 
&  \lambda_{j}(\Y_U) - \lambda_{n}(\Y_U) \le 0.002 \kappa \lammin  < 0.25 \lammin.
\end{align*}
Thus, w.p. $\ge 1 - p_{U,1}$, $j=r$ is the smallest index for which $\lambda_r(\Y_U) - \lambda_{n}(\Y_U) \ge 0.25 \lammin$ and hence the rank estimation approach of Corollary \ref{mainthm_unknown_r_2} returns $\hat{r}=r$.

\section{Conclusions and Future Work} \label{conclusions}

We presented two iterative phase retrieval algorithms -- LRPR1 and LRPR2 -- for recovering a set of $q$ unknown vectors lying in a low ($r$) dimensional subspace of $\Re^n$ from their phaseless measurements. 
Both methods are initialized by a two step spectral initialization procedure, called LRPR-init, that first estimates the subspace from which all the vectors are generated, and then estimates the projection, of each vector, into the estimated subspace.
The rest of LRPR1 involves projected truncated gradient descent. The remainder of LRPR2 involves alternating minimization to update the estimates of $\U$, $\B$, and the unknown phase of $(\a_{i,k}{}'\x_k)$ for each $i,k$.

We obtained sample complexity bounds for LRPR-init and argued that, when $q/r$ is large, these are much smaller than those for TWF or any other single-vector PR method. 
Via extensive experiments, we also showed that the same is true for both the complete algorithms - LRPR1 and LRPR2. Between the two, LRPR2 has better performance, but also higher per iteration computational cost, than LRPR1.

In future work we will analyze the complete LRPR2 algorithm. This should replace the dependence of sample complexity on $1/\varepsilon^2$ by a dependence on $-\log \varepsilon$. 
%

\bibliographystyle{IEEEbib}
\bibliography{../../bib/tipnewpfmt_kfcsfullpap}

\appendices
\renewcommand{\thetheorem}{\thesection.\arabic{theorem}}

\section{Preliminaries} \label{prelim}


As explained in \cite{vershynin}, $\epsilon$-nets are a convenient means to discretize compact metric spaces. The following definition is \cite[Definition 5.1]{vershynin} for the unit sphere.
\begin{definition}[$\epsilon$-net and covering number of the unit sphere in $\Re^n$]
For an $\epsilon > 0$, a subset $\N_\epsilon$ of the unit sphere in $\Re^n$ is called an $\epsilon$-net if, for every vector $\x$ on the unit sphere, there exists a vector $\y \in \N_\epsilon$ such that $\|\y - \x\| \le \epsilon$.

The covering number of  the unit sphere in $\Re^n$, is the size of the smallest $\epsilon$-net, $\N_\epsilon$, on it.
\end{definition}

\begin{fact}[Facts about $\epsilon$-nets] \em \label{fact_nets}
\ 
\begin{enumerate}
\item By Lemma 5.2 of \cite{vershynin}, the covering number of the unit sphere in $\Re^n$ is upper bounded by $(1+\frac{2}{\epsilon})^n$. 

\item By Lemma 5.4 of \cite{vershynin}, for a symmetric matrix, $\W$,
$
\|\W\| = \max_{\x: \|\x\|=1} \|\x' \W \x\| \le \frac{1}{1-2 \epsilon} \max_{\x \in \N_{\epsilon} } \|\x' \W \x\|.
$
\end{enumerate}
\end{fact}

%
%

\begin{fact}[Facts about sub-Gaussian random vectors] \em \label{fact_versh} \
\begin{enumerate}
\item  \label{sub_expo}
If $\x$ is a sub-Gaussian random vector with sub-Gaussian norm $K$, then for any vector $\z$, (i) $\x'\z$ is sub-Gaussian with sub-Gaussian norm bounded by $K \|\z\|$; (ii) $(\x'\z)^2$ is sub-exponential with sub-exponential norm bounded by $2 K^2 \|\z\|^2$; and (iii) $(\x'\z)^2 - \E[(\x'\z)^2]$ is centered (zero-mean), sub-exponential with sub-exponential norm bounded by  $4 K^2 \|\z\|^2$. This follows from the definition of a sub-Gaussian random vector; Lemma 5.14 and Remark 5.18 of \cite{vershynin}.

\item \label{cor_5_17}
By \cite[Corollary 5.17]{vershynin}, if $x_i$, $i=1,2, \dots N$, are a set of independent, centered, sub-exponential r.v.'s with sub-exponential norm bounded by $K_e$, then,
for an $\varepsilon < 1$,
\[
\Pr\left( |  \sum_{i=1}^N x_i| > \varepsilon K_e N \right) \le 2 \exp( - c \varepsilon^2  N).
\]

\item \label{K_gauss_vec}
If $\x \sim \n(0, \Lambar)$ with $\Lambar$ diagonal, then $\x$ is sub-Gaussian with $\|\x\|_{\varphi_2} \le c \sqrt\lammax$.
Moreover, if  $\y = x_1 \x$ where $x_1$ is a zero mean bounded r.v. with bound $M$, then $\|\y\|_{\varphi_2} \le c M \sqrt{\lammax}$.




\item \label{rayleigh_type_bnd}
If $\x_i \sim \n(0, \Lambar)$, for $i=1,2,\dots, N$, are $n$-length random vectors and  $\Lambar$ is diagonal, then
\begin{align*}
& \Pr \left( \max_{i=1,2,\dots, N} \left\|\x_i \right\|^2 \le \lammax \cdot n \cdot 2 \nu \right)   \\
& \ge 1 - 2n N \exp(-\nu) , \ \text{for} \   \nu > 1.
\end{align*}
This is a direct consequence of eq. 5.5 of \cite{vershynin} which says that if $x \sim \n(0,1)$, then $\Pr(|x_i| > t) \le 2 \exp(-t^2/2)$ for a $t>1$.
Using this along with the union bound first for bounding $\|\x_i\|^2 = \sum_{j=1}^n (\x_i)_j^2$ for a given $i$ and then for bounding its $\max$ over $i$ gives the above result.

\item \label{subG_bnd}
Using \cite[Lemma 5.5]{vershynin}, if $\x_i$'s are sub-Gaussian random vectors with sub-Gaussian norm bounded by $K$, then the following generalization of the above fact holds:
$
\Pr \left( \max_{i=1,2,\dots, N} \|\x_i\|^2 \le K^2 \cdot n \cdot 2 \nu \right)  \ge 1 - Cn N \exp(- c\nu).
$


\end{enumerate}
\end{fact}

The following is an easy corollary of Cauchy-Schwartz for sums of products of vectors. 
\begin{theorem}[Cauchy-Schwartz for sums of matrices]
\label{CSmat}
For matrices $\bm{X}_t$ and $\bm{Y}_t$,
$
\left\|\frac{1}{\alpha} \sum_{t=1}^{\alpha} \tilde{\bm{X}}_t {\tilde{\bm{Y}}_t}'\right\|^2 \leq \lambda_{\max}\left(\frac{1}{\alpha} \sum_{t=1}^{\alpha} \tilde{\bm{X}}_t {\tilde{\bm{X}_t}}'\right)
\lambda_{\max}\left(\frac{1}{\alpha} \sum_{t=1}^{\alpha} \tilde{\bm{Y}}_t {\tilde{\bm{Y}}_t}'\right).
$
\end{theorem}

\section{Proofs of lemmas from Section \ref{outline}}
\label{proof_lems}
We prove the lemmas that were not proved in Section \ref{outline}.

Define
\[
\w_{i,k} := \left( \a_{i,k}{}'\frac{\x_k}{\|\x_k\|} \right) \a_{i,k} \one_{ \left(\a_{i,k}{}'\frac{\x_k}{\|\x_k\|} \right)^2 \le 9 (1-\epsilon_1)} \ \|\x_k\|.
\]
Then $\Y^- = \frac{1}{mq} \sum_k \sum_i \w_{i,k} \w_{i,k}{}'$.

\begin{proof}[Proof of Lemma \ref{expec_Y}]
Since $\a_{i,k}$ is rotationally symmetric, to compute $\E[\w_{i,k} \w_{i,k}{}' | \X]$ easily, we can let $\frac{\x_k}{\|\x_k\|}$ be the first column of the identity matrix. With this,
$\w_{i,k}  = (\a_{i,k})_1 \a_{i,k} \one_{ (\a_{i,k})_1^2 \le 9 (1-\epsilon_1)} \ \|\x_k\|.$ Thus, using the argument 
of \cite[Appendix C]{twf},
$\E[\w_{i,k} \w_{i,k}{}' | \X] = (\beta_1^-\frac{\x_k}{\|\x_k\|} \frac{\x_k}{\|\x_k\|}' + \beta_2^-\I ) \|\x_k\|^2 = (\beta_1^- \x_k \x_k{}' + \beta_2^- \|\x_k\|^2 \I ) $.
Using this with $\x_k = \U \b_k$, $\Lambar= \frac{1}{q} \sum_k \b_k \b_k{}'$, and $\frac{1}{q}\sum_k \|\b_k\|^2 =  \trace(\Lambar)$, both claims follow under Model \ref{det_mod}.
%
%
To get $\bm\Sigma^-$ under Model \ref{gauss_mod}, using linearity of expectation and of trace, $\trace(\Lambar) =  \E[\frac{1}{q} \sum_k \|\b_k\|^2]$.
\end{proof}

\begin{proof}[Proof of Lemma \ref{Y_minus}]
The proof relies on the following.
\begin{itemize}
\item  Let $D = \max_k \|\b_k\|^2  =\max_k \|\x_k\|^2$.
\item We first argue argue that, conditioned on $\X$, each $\w_{i,k}$ is sub-Gaussian with sub-Gaussian norm bounded by $c \|\x_k\| \le c\sqrt{D}$.

To show this easily, we use the strategy of \cite[Appendix C]{twf}. Since $\a_{i,k}$ is rotationally symmetric, without loss of generality, suppose that $\frac{\x_k}{\|\x_k\|}$ is the first column of the identity matrix. Then, $\w_{i,k}  = \a_{i,k}  (\a_{i,k})_1 \one_{ (\a_{i,k})_1^2 \le 9 (1-\epsilon_1)} \ \|\x_k\|$.
With this simplification, $\w_{i,k}$ is of the form $\x x_1$ where $x_1$ is a bounded r.v. with bound $\sqrt{9 (1-\epsilon_1)}\|\x_k\|$ and $\x$ is Gaussian with zero mean and covariance matrix $\I$. Thus, using Fact \ref{fact_versh}, item \ref{K_gauss_vec}, it is sub-Gaussian with sub-Gaussian norm bounded by $c \|\x_k\| \le c \sqrt{D}$.


\item  Conditioned on $\X$, all the $\w_{i,k}$'s are mutually independent. There are $N=mq$ of them.

\end{itemize}
Thus, all the $mq$ $\w_{i,k}$'s are  mutually independent sub-Gaussian random vectors with sub-Gaussian norm bounded by $c\sqrt{D}$. So, we can apply the second claim of Theorem \ref{versh} with $\w_j$ replaced by $\w_{i,k}$ and summed over the $N=mq$ vectors, $\w_{i,k}$, to show that $\|\Y^- - \E[\Y^-| \X]\| \le \epsilon_2 D$ w.p. $\ge 1 - 2\exp(n \log 9 -c \epsilon_2^2 m q )$.
%
\end{proof}

\begin{proof}[Proof of Lemma \ref{D_bnd}]
Let $D = \max_k \|\b_k\|^2$.
Under Model \ref{det_mod}, $D \le r \rho \lammax$ by definition.
Under Model \ref{gauss_mod}, we use Fact \ref{fact_versh}, item \ref{rayleigh_type_bnd} with $n \equiv r$, $N \equiv q$, $\nu \equiv 5 \log n$ to get $D \le 10\log n$ w.p. $\ge 1 - 2 r q n^{-5} \ge 1 - 2 q n^{-4}$ since $r \le n$.
\end{proof}

\begin{proof}[Proof of Lemma \ref{EY_minus}]
$\| \E[\Y^-| \X] - \bm\Sigma^- \|$ is bounded by
\begin{align*}
\beta_1^- \| \frac{1}{q} \sum_k \b_k \b_k{}' - \Lambar \|  +
                                      \beta_2^- \left|\frac{1}{q} \sum_k \| \b_k \|^2 - \mathrm{trace}(\Lambar) \right|
\end{align*}
Clearly, $\beta_1^- \le 2$ and $\beta_2^- \le 1$.
By Fact \ref{fact_versh}, item \ref{K_gauss_vec}, $\b_k$ is sub-Gaussian with $\|\b_k\|_{\varphi_2} \le c \sqrt{\lambda_{k,\max}}$. By model assumption,  $\lambda_{k,\max} \le c \lammax$. Thus, $\|\b_k\|_{\varphi_2} \le c \sqrt{\lammax}$. Apply the second claim of Theorem \ref{versh} with $N \equiv q$, $n \equiv r$ and $K \equiv c \sqrt\lammax$ to get $\left\|\frac{1}{q} \sum_k \b_k \b_k' - \Lambar \right\| \le \epsilon_3 \lammax$  w.p. $ \ge 1 - 2\exp(r \log 9 - c\epsilon_3^2 q)$. The constant next to $\epsilon_3 \lammax$ has been absorbed into the $c$ in the probability.
\\
For the second term, apply  Theorem \ref{versh} with $N \equiv r q$, $n \equiv 1$ and $K \equiv c \sqrt\lammax$ to get
$\left|\frac{1}{rq} \sum_k \| \b_k \|^2 - \frac{\mathrm{trace}(\Lambar)}{r} \right| \le  \epsilon \lammax$ w.p. $ \ge 1 - 2\exp(\log 9 - c\epsilon^2 rq)$. Thus,  $\left|\frac{1}{q} \sum_k \| \b_k \|^2 - \mathrm{trace}(\Lambar) \right| \le  r \epsilon \lammax$ with the same probability. Use $\epsilon = \epsilon_3/r$ to conclude that $\left|\frac{1}{q} \sum_k \| \b_k \|^2 - \mathrm{trace}(\Lambar) \right| \le \epsilon_3 \lammax$ w.p. $ \ge 1 - 18 \exp(- c\epsilon_3^2 \frac{q}{r})$.%
\end{proof}

\begin{proof}[Proof of Lemma \ref{Sigma_minus}]

Using $\xi \exp(-\xi^2/2 ) < 1 $ for all $\xi^2 > 8$,
\begin{align*}
\beta_2^+(\epsilon_1) - \beta_2^-(\epsilon_1) & = \E[\xi^2 \one_{ (9-9\epsilon_1 ) \le \xi^2 \le (9 + 9\epsilon_1 ) }] \\
&  = 2 \int_{\sqrt{9-9\epsilon_1}}^{\sqrt{9+9\epsilon_1}}  \xi^2 \frac{1}{\sqrt{2\pi}}  e^{-\frac{\xi^2}{2}} d \xi  \\
& \le 2 \frac{1}{\sqrt{2\pi}} \int_{\sqrt{9-9\epsilon_1}}^{\sqrt{9+9\epsilon_1}}  \xi d\xi =  \frac{1}{\sqrt{2\pi}} (18 \epsilon_1) 
\end{align*}
Similarly, using $\xi^3 \exp(-\xi^2/2 ) < 3.1$ for all $\xi^2 > 8$,
\begin{align*}
\beta_1^+(\epsilon_1) - \beta_1^-(\epsilon_1)
&\le  3.1 \cdot \frac{1}{\sqrt{2\pi}}  18 \epsilon_1 < 22.4 \epsilon_1
\end{align*}
\end{proof}

\begin{proof}[Proof of Lemma \ref{beta1_minus_bnd}]
Using $(\xi^4 - \xi^2) \exp(-\xi^2/4 ) < 7.58$ for all $\xi^2 > 8$, Fact \ref{fact_versh}, item \ref{rayleigh_type_bnd} and $\epsilon_1 \le 1/9$,
\begin{align*}
\beta_1^- & = \E[(\xi^4 - \xi^2)] -  \E[(\xi^4 - \xi^2) \one_{\xi^2 \ge (9-9\epsilon_1)}] \\
& \ge 2 - 7.58 \cdot 2 \int_{\sqrt{9-9\epsilon_1}}^\infty  \frac{1}{\sqrt{2\pi}} e^{-\frac{\xi^2}{4}} d \xi  \\
& = 2 -  7.58 \cdot  \sqrt{2} \Pr(x^2 \ge (9-9\epsilon_1)/2 )   \\
& \ge 2 -  7.58 \cdot \sqrt{2} \exp(-(9-9\epsilon_1)/4) \ge 0.5
\end{align*}
where $x \sim \n(0,1)$. 
%
\end{proof}

\begin{proof}[Proof of Lemma \ref{ng_bnd_lem}]
By triangle inequality, $| \hatng - \ng | \le |\hatng - \|\x\| | + | \|\x\|- \ng \ |$. By Fact \ref{ng_bnd}, $| \|\x\| - \ng | \le \delta_U \|\b\|$.
We can bound $|\hatng - \|\x\| |$ by applying Theorem \ref{versh}, part 1. Recall that $\hatng^2  = \frac{1}{\tildm} \sum_i \y_i =\frac{1}{\tildm} \sum_i (\a_i{}' \x)^2 $ and so, w.p. $\ge 1- 2\exp(-c\epsilon_4^2 \tildm)$,
\[
| \ \hatng^2 - \| \x\|^2 \ | = | \x' (\frac{1}{m} \sum_i \a_i \a_i{}' - \I) \x |  \le \epsilon_4 \|\x\|^2
\]
Since $\epsilon_4<1$, this implies that $(1-\epsilon_4)\| \x\|  \le \hatng \le (1+\epsilon_4) \| \x\|$ and so
 $| \hatng - \|\x\| \ | \le \epsilon_4 \|\x\|$.
Hence, w.p. $\ge 1- 2\exp(-c\epsilon_4^2 \tildm )$, $| \hatng - \ng | \le (\epsilon_4 + \delta_U)\|\x\|$.
\end{proof}

\begin{proof}[Proof of Lemma \ref{Yg_bnd}] The proof is a simplified and clearer version of the proof of Theorem 4.1 of \cite{pr_altmin}. The complete proof is given in the Supplementary Document. We give the key ideas here.
Let $\E[.]$ denote expectation conditioned on $\x, \Uhat$. Recall that $\E[\Y_g] = 2 \g \g' + \|\g\|^2 \I$ and thus we need to bound $\|\Y_g - \E[\Y_g]\|$.
 (1) Due to rotational symmetry of $\ta_i$'s we can let $\g$ be the first column of $\I$. This gives a simpler expression for $\Y_g$.
(2) Truncate $\ta_i$'s as follows: for each $j=1,2,\dots,r$, let $(\tta_i)_j = (\ta_i)_j \ \text{if} \  ((\ta_i)_j )^2 \le 20\log m, \ \ (\tta_i)_j = 0 \ \text{otherwise}$. Define $\Y_g^{\mathrm{trunc}}$ using $\tta_i$'s.
(3) Apply Theorem 1.4 of \cite{tail_bound} (matrix Bernstein) to bound $\|\Y_g^{\mathrm{trunc}} - \E[\Y_g^{\mathrm{trunc}}] \|$ w.p. $\ge 1- 2/\tilde{m}^8$.
(4) By definition, $\|\Y_g - \Y_g^{\mathrm{trunc}} \|=0$ w.p. $\ge 1- 2/\tilde{m}^8$.
(5) Finally bound $\|\E[\Y_g] - \E[\Y_g^{\mathrm{trunc}}] \|$ by  $4/\tilde{m}^{4.5}$. This is easy because both $\E[\Y_g]$ and $ \E[\Y_g^{\mathrm{trunc}}]$ are diagonal (the latter is diagonal because the truncation ensures that entries of $\tta_i$ are also mutually independent and zero mean). Bound the diagonal entries using the following trick: for $\xi$ large, e.g. for $\xi > 10$, $\xi^4 e^{-\xi^2/4} < 1$ and so $\xi^4 e^{-\xi^2/2} < e^{-\xi^2/4}$; similarly, $\xi_1^2 e^{-\xi_1^2/4} \xi_2^2 e^{-\xi_2^2/4} < 1$ for $\xi_1,\xi_2$ large.
\end{proof}



\clearpage

\section{Supplementary Document} \label{proof_versh}


\begin{proof}[Proof of Theorem \ref{versh}]
%
The proof strategy is similar to that of Theorem 5.39 of \cite{vershynin}.
By Fact \ref{fact_versh}, item \ref{sub_expo}, for each $j$, the r.v.s $\w_j{}' \z$ are sub-Gaussian with sub-Gaussian norm bounded by $K \| \z\|$;  $(\w_j{}' \z)^2$ are sub-exponential with sub-exponential norm bounded by $2K^2 \| \z\|^2$; and $(\w_j{}' \z)^2 - \E[(\w_j{}' \z)^2] = \z'(\w_j \w_j{}') \z - \z'(\E[\w_j \w_j{}']) \z$ are centered sub-exponential with sub-exponential norm bounded by $4K^2 \| \z\|^2$. Also, for different $j$'s, these are clearly mutually independent. Thus, by applying Fact \ref{fact_versh}, item \ref{cor_5_17} (Corollary 5.17 of \cite{vershynin}) we get the first part.

To prove the second part, let $\N_{1/4}$ denote a 1/4-th net on the unit sphere in $\Re^n$. Let $\W:= \frac{1}{N} \sum_{j=1}^N ( \w_j \w_j{}' - \E[ \w_j \w_j{}'])$.
Then by Fact \ref{fact_nets} (Lemma 5.4 of \cite{vershynin})
\beq \label{epsnet_bnd}
\|\W\| \le 2 \max_{\z \in  \N_{1/4}} |\z' \W \z|
\eeq
Since $\N_{1/4}$ is a finite set of vectors, all we need to do now is to bound $|\z' \W \z|$ for a given vector $\z$ followed by applying the union bound to bound its  maximum over all ${\z \in  \N_{1/4}}$.
The former has already been done in the first part.
By Fact \ref{fact_nets} (Lemma 5.2 of \cite{vershynin}), the cardinality of $\N_{1/4}$ is at most $9^n$. Thus, using the first part, 
$
\Pr\left(\max_{\z \in  \N_{1/4}} |\z' \W \z|  \ge  \frac{4\varepsilon K^2}{2} \right) \le 9^n \cdot 2 \exp(-c \frac{\varepsilon^2}{4} N) =  2 \exp(n \log 9 -c \varepsilon^2 N).
$
By \eqref{epsnet_bnd}, we get the result.
\end{proof}


\begin{proof}[Complete Proof of Lemma \ref{Yg_bnd}] The proof is a simplified and clearer version of the proof of Theorem 4.1 of \cite{pr_altmin}.
The few differences are as follows:  we truncate differently (in a simpler fashion); and we use different constants to get a higher probability of the good event. 

Without loss of generality, assume that $\g$ is unit norm.
Recall that $\Y_g:=\frac{1}{m} \sum_i (\ta_i{}'\g)^2  \ta_i \ta_i{}'$ with $\ta_i \iidsim \n(0,\I)$. Since $\ta_i$'s are rotationally symmetric, without loss of generality, we can assume that $\g$ is the first column of identity matrix. Then,
\[
\Y_g = \frac{1}{m} \sum_i (\ta_i)_1^2  \ta_i \ta_i{}'
\]
We use Theorem 1.5 of \cite{tail_bound} to prove the result. To use this result, we need to first truncate $\ta_i$'s. In particular we need to definitely truncate $(\ta_i)_1$ and $\|\ta_i\|^2$.
However truncating all entries of $\ta_i$ results in a simpler proof and hence we use this approach.
For $j=1,2,\dots, r$, define
\begin{align*}
& (\tta_i)_j = (\ta_i)_j \ \text{if} \  ((\ta_i)_j )^2 \le 20 \log m, \\
& (\tta_i)_j = 0 \ \text{otherwise}
\end{align*}
Define
\[
\Y_g^{\mathrm{trunc}}:= \frac{1}{\tilde{m}} \sum_i (\tta_i)_1^2  \tta_i \tta_i{}'
\]
By Fact \ref{fact_versh}, item \ref{rayleigh_type_bnd},  $\tta_i = \ta_i$ w.p. $\ge 1 - 2r / \tilde{m}^{10} \ge 1 - 2/\tilde{m}^9$ since $\tilde{m} \ge r$. This holds for all $i=1,2,\dots \tildm$, w.p. $\ge 1 - 2/\tilde{m}^8$. Thus, $\|\Y_g - \Y_g^{\mathrm{trunc}}\|=0$ w.p. $\ge 1 - 2/\tilde{m}^8$.

To apply Theorem 1.5 of \cite{tail_bound} (matrix Bernstein), define matrix $X_i: = (\tta_i)_1^2  \tta_i \tta_i{}'$. Clearly $\|X_i\| \le 20 \log \tilde{m} \cdot (r 20 \log \tilde{m}) = 400 r \log^2 \tilde{m} :=R$.
Also $\|\E[X_i^2 ]\| \le  (\tta_i)_1^4 \|\tta_i\|^2  \| \E[\tta_i \tta_i{}']\| \le  8000 r \log^3 \tilde{m} \|\E[\tta_i \tta_i{}']\| \le 8000 r \log^3 \tilde{m} $. Here we used $\|\E[\tta_i \tta_i{}']\| \le 1$. This is true because: (a) even with the truncation, the different components of $\tta_i$ remain independent and zero mean and so $\E[(\tta_i)_{j1} (\tta_i)_{j2}] = 0$ for $j1 \neq j2$; thus, $\E[\tta_i \tta_i{}']$ is diagonal;  and (b) it is easy to see that $\E[(\tta_i)_{j}^2] \le \E[(\ta_i)_{j}^2] = 1$.

Thus, we can apply the theorem with $R = 400 r \log^2 \tilde{m}$ and $\sigma^2=\|\sum_i \E[X_i^2 ]\| \le 20 \tildm R \log \tildm$. Picking $\nu =  \sqrt{400 \cdot 20 \frac{r \log^4 \tilde{m}}{\tilde{m}}}$, we get
\[
\|\Y_g^{\mathrm{trunc}} - \E[\Y_g^{\mathrm{trunc}}] \| \ge \nu
\]

\begin{align*}
\text{w.p.} & \le 2r\exp\left(- \frac{\tilde{m}^2 \nu^2}{\tilde{m}R \log \tildm + \tilde{m}R \nu/3} \right) \\
& \le 2r\exp\left(- \frac{\tilde{m} \nu^2}{2R \log \tildm } \right) \\
& \le 2 \exp\left(\log r - \frac{\tilde{m} 400 \cdot 20 r \log^4 \tilde{m}}{\tilde{m} 2 \cdot 400 r \log^3 \tilde{m}}\right) \\
&  = 2 \exp(\log r - 10 \log \tilde{m}) \le 2/\tilde{m}^9
\end{align*}
This follows since $\tilde{m} > c r \log^4 r $ (and so $\nu < 1$ and $\tilde{m} > r$) 

Moreover,  w.p. $\ge 1 - 2/\tilde{m}^8$,
\[
\|\Y_g^{\mathrm{trunc}} - \Y_g\| = 0
\]
Thus, w.p. $\ge 1 - 4/\tilde{m}^8$,
\[
\|\Y_{\g} -  \E[\Y_g^{\mathrm{trunc}}] \|  \le \nu =  \sqrt{8000 \frac{r \log^4 \tilde{m}}{\tilde{m}}} 
\]
Now we only need to bound $\|\E[\Y_g^{\mathrm{trunc}}] - \E[\Y_{\g}]\|$. This is easy and uses the following facts.
(a) clearly, $\E[\Y_{\g}]$ is diagonal; (b) $\E[\Y_g^{\mathrm{trunc}}]$ is also diagonal since with our truncation the different components of $\tta_i$ remain independent and zero mean; and (c) thus we only need to bound the diagonal entries of $\E[\Y_g^{\mathrm{trunc}}] - \E[\Y_{\g}]$.
Consider the (1,1)-th entry. We bound this by using the fact that for $\xi > 10$, $\xi^4 e^{-\xi^2/4}  < 1 $. If $\tilde{m} > 3$, $20\log \tilde{m} > 18 > 10$ and hence this bound holds over the entire region of integration.
\begin{align*}
\E[ (\ta_i)_1^4 - (\tta_i)_1^4] &= 2\int_{\sqrt{20\log \tilde{m}}}^\infty \xi^4 \frac{ e^{-\xi^2/2} }{\sqrt{2\pi}} d \xi  \\
& \le 2\int_{\sqrt{20\log \tilde{m}}}^\infty  \frac{ e^{-\xi^2/4} }{\sqrt{2\pi \cdot 2}} \sqrt{2} d \xi  \\
& = \sqrt{2} \Pr(|x| > \sqrt{9\log \tilde{m}/2}) \le \sqrt{2} \frac{2}{\tilde{m}^{4.5}}
\end{align*}
if $\tilde{m} > 3$. Here $x$ is a standard Gaussian r.v.. The last inequality used Fact \ref{fact_versh}, item \ref{rayleigh_type_bnd}.

Next consider the $(j,j)$-th entry for $j>1$. This can be bounded using a similar trick. 
\begin{align*}
& \E[(\ta_i)_1^2 (\ta_i)_j^2  - (\tta_i)_1^2 (\tta_i)_j^2 ] \\
&\le  2 \int_{\xi_1^2 \ge 20\log \tilde{m}} \xi_1^2 \xi_2^2 \exp(-(\xi_1^2 + \xi_2^2)/2) \frac{1}{{2\pi}} d \xi_1 d \xi_2   \\
&=  2 \int_{\xi_1^2 \ge 20\log \tilde{m}} \xi_1^2 \exp(-\xi_1^2/2) \frac{1}{\sqrt{2\pi}} d \xi_1  \\
& \le  2\int_{\xi_1^2 \ge 20\log \tilde{m}} \exp(-\xi_1^2/4) \frac{\sqrt{2}}{\sqrt{2\pi 2}} d \xi_1
\le  \frac{2\sqrt{2}}{\tilde{m}^{4.5}}
\end{align*}
Thus,  w.p. $\ge 1 - 4/\tilde{m}^8$,
\[
\|\Y_{\g} - \E[\Y_{\g}]\| \le \sqrt{8000 \frac{r \log^4 \tilde{m}}{\tilde{m}}} +  \frac{4}{\tilde{m}^{4.5}}.  
\]
If $\tilde{m} > c r\log^4 r / \epsilon_5^2$, then the above bound is below $ 2 \epsilon_5$. To see this notice that for $\tilde{m} = c r\log^4 r / \epsilon_5^2$, $\frac{\log^4 \tilde{m}}{\tilde{m}} \le c \epsilon_5^2 / r$; and  $(\log^4 \tilde{m})/\tilde{m}$ is an increasing function (for $\tilde{m}$ large).
\end{proof}

\input{plane_expt_describe}

\end{document}

%% file: plane.tex

\begin{figure}[t]
	\centering{
\scalebox{0.65}{
	\begin{tabular}{cc}
\includegraphics[height=1.5cm,trim=1.5cm 1.5cm 1.5cm 1.5cm,clip]{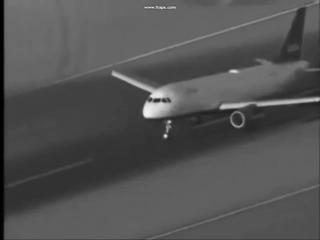}\hspace{0.2cm}
\includegraphics[height=1.5cm,trim=1.5cm 1.5cm 1.5cm 1.5cm,clip]{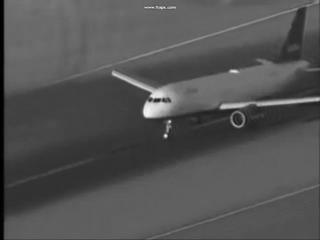}\hspace{0.2cm}
\includegraphics[height=1.5cm,trim=1.5cm 1.5cm 1.5cm 1.5cm,clip]{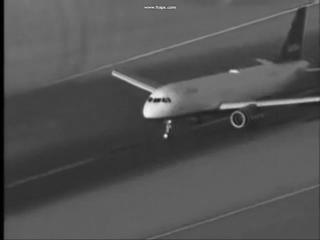}\hspace{0.2cm}
\includegraphics[height=1.5cm,trim=1.5cm 1.5cm 1.5cm 1.5cm,clip]{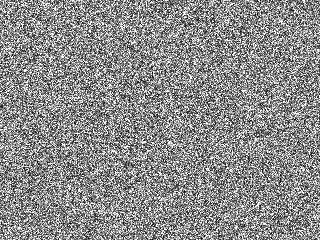}\hspace{0.2cm}
\includegraphics[height=1.5cm,trim=1.5cm 1.5cm 1.5cm 1.5cm,clip]{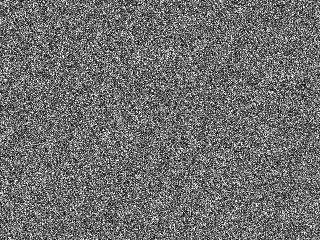}\\
\includegraphics[height=1.5cm,trim=1.5cm 1.5cm 1.5cm 1.5cm,clip]{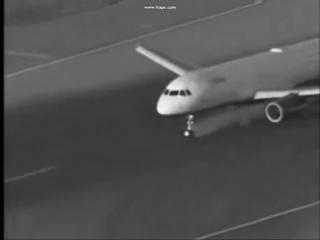}\hspace{0.2cm}
\includegraphics[height=1.5cm,trim=1.5cm 1.5cm 1.5cm 1.5cm,clip]{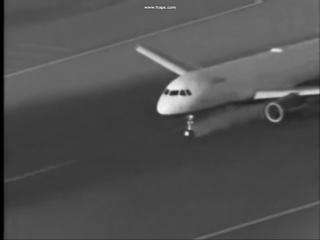}\hspace{0.2cm}
\includegraphics[height=1.5cm,trim=1.5cm 1.5cm 1.5cm 1.5cm,clip]{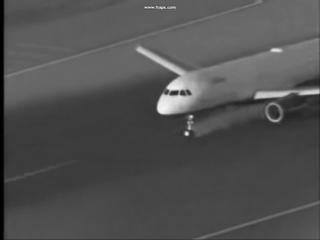}\hspace{0.2cm}
\includegraphics[height=1.5cm,trim=1.5cm 1.5cm 1.5cm 1.5cm,clip]{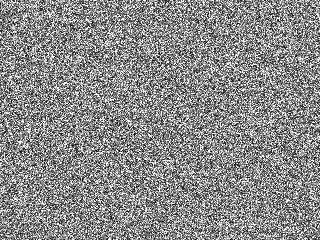}\hspace{0.2cm}
\includegraphics[height=1.5cm,trim=1.5cm 1.5cm 1.5cm 1.5cm,clip]{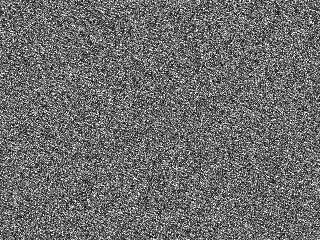}\\
    	\hspace*{\fill}\makebox[0pt]{Original}\hspace*{\fill}
		\hspace*{\fill}\makebox[0pt]{LRPR2}\hspace*{\fill}
		\hspace*{\fill}\makebox[0pt]{LRPR1}\hspace*{\fill}
		\hspace*{\fill}\makebox[0pt]{basic TWFproj}\hspace*{\fill}
		\hspace*{\fill}\makebox[0pt]{basic TWF}\hspace*{\fill}\\
	\end{tabular}
}
}
\vspace{-0.1in} 
\caption{\small\sl{First column: frame 1 and 104, of the original plane video. Next three columns: frames recovered using the various methods from $m=3n$ phaseless masked Fourier (CDP model) measurements. 
}}
\vspace{-0.1in}
\label{planefig}
\end{figure}

%% file: new_new_expts.tex
\begin{table*}[t!]
\centering{
\begin{tabular}{|c|c|c|c|c|c|c|c|c|}
\hline
$m/n$ & $\Pr(\rhat=r)$ & LRPR ($r$ known) & LRPR ($r$ unknown)  & LRPR ($\rhat=2r$) & LRPR (noisy meas) & TWFproj & TWF  & LRPR-same\\
\hline 
 \multicolumn{9}{|c|}{$\tot = 100$}  \\
\hline
0.10&		0.24&				1.23&		1.31&		1.22&		1.37		& 0.96&		1.63  &  0.99
	\\
\hline
0.50&		0.68&				0.33&		0.45&		0.55&		0.52		& 0.67&		1.47  &  1.03 \\
\hline
1.00&		1.00&				0.17&		0.17&		0.30&		0.19		& 0.53&		1.34  & 1.04 	\\
\hline
 \multicolumn{9}{|c|}{$\tot = 1000$}  \\
\hline
0.10&		1.00&				0.39&		0.39&		0.70&		0.47		& 0.71&		1.62 & 0.98 	\\
\hline
0.50&		1.00&				0.10&		0.10&		0.26&		0.12		& 0.57&		1.47	& 1.02 \\
\hline
1.00&		1.00&				0.05&		0.05&		0.13&		0.06		& 0.48&		1.34 & 1.04 \\
\hline
\end{tabular}
}
\caption{\small\sl{Real Gaussian measurement vectors: initialization error comparisons. LRPR: LRPR-init, TWF: TWF-init, TWFproj: TWFproj-init.}}
\label{real_gauss}
\end{table*}

\begin{table*}[t!]
\centering{
\begin{tabular}{|c|c|c|c|c|c|c|c|}
\hline
$m/n$ & $\Pr(\rhat=r)$ & LRPR ($r$ known) & LRPR ($r$ unknown)  & LRPR ($\rhat=2r$) & LRPR (noisy meas) & TWFproj & TWF \\ 
\hline 
 \multicolumn{8}{|c|}{$\tot = 100$}  \\
\hline
0.10&		0.21&				1.29&		1.32&		1.30&		1.37		&  0.97&		1.66	\\
\hline
0.50&		0.41&				0.53&		0.67&		0.76&		0.75		&  0.76&		1.55\\
\hline
1.00&		0.97&				0.27&		0.29&		0.47&		0.36		&  0.63&		1.45	\\
\hline
7.00&		1.00&				0.04&		0.04&		0.07&		0.05	    &   0.13&		0.65	\\
\hline
 \multicolumn{8}{|c|}{$\tot = 1000$}  \\
\hline
0.10&		0.99&				0.51&		0.51&		0.84&		0.62	&	   0.75&		1.67	\\
\hline
0.50&		1.00&				0.15&		0.15&		0.38&		0.19	&	   0.63&		1.54	\\
\hline
1.00&		1.00&				0.08&		0.08&		0.21&		0.10	&	   0.55&		1.45	\\
\hline
7.00&		1.00&				0.01&		0.01&		0.03&		0.02&  	0.11&		0.65	\\
\hline
\end{tabular}
}
\caption{\small\sl{Complex Gaussian measurement vectors: initialization comparisons. LRPR: LRPR-init, TWF: TWF-init, TWFproj: TWFproj-init.}}
\label{complex_gauss}
\end{table*}

We show the power of both LRPR1 and LRPR2 for recovering a real video from coded diffraction pattern (CDP) measurements in Fig. \ref{planefig}. As can be seen, with as few as $m=3n$ CDP measurements, both these methods significantly outperform basic TWFproj (Algorithm \ref{twfproj_iter_algo} initialized using Algorithm \ref{twfproj_init_algo}) and basic TWF (Algorithm \ref{twf_iter_algo} initialized using Algorithm \ref{twf_init_algo}). This experiment is inspired by an analogous experiment for recovering a regular camera image from CDP measurements reported in \cite[Fig. 2]{twf}.
While this is not a real practical application since the video used is a regular camera video of a moving airplane, this example illustrates two points: (i) many real image sequences are indeed approximately low-rank; and (ii) our algorithm has significant advantage over single vector PR methods for jointly recovering this approximately low-rank video. For a detailed explanation of this and some more such experiments, please see Supplementary Material and \url{http://www.ece.iastate.edu/~namrata/LRPR/}.

\section{Numerical Experiments} \label{sims}
We discuss here the results of three sets of experiments. All experiments were done on a single laptop which had these specifications: Intel(R) CPU E3-1240 v5 \@ 3.50 GHz, Installed memory: 32 GB, System type is 64 bit.

{\bf Experiment 1. }
The first experiment shows the power of the proposed initialization approach, LRPR-init (Algorithm~\ref{algo_init}), by comparing its initialization error with that of TWF initialization (TWF-init, Algorithm \ref{twf_init_algo}) and of TWFproj-init (Algorithm \ref{twfproj_init_algo}). TWF-init does not use knowledge of rank, TWFproj-init assumes $r$ is known,  while LRPR-init estimates the rank automatically as explained earlier. For a fair comparison with TWFproj-init, we also show the error of  LRPR-init with $\hat{r}=r$. Data was generated as follows. The matrix $\U$ is obtained by orthonormalizing an $n \times r$ matrix with iid Gaussian entries; $\b_k$'s were generated as being iid uniformly distributed between $-1$ and $1$; and we set $\x_k = \U \b_k$. Measurements were generating using \eqref{exact_mod}.

\begin{algorithm}
\caption{TWF initialization (TWF-init)}
\label{twf_init_algo}
For each $k=1,2,\dots,q$, set $\xhat_k^0$ as the top eigenvector of $\sum_{i=1}^m \y_{i,k} \a_{i,k} \a_{i,k}{}'\mathbbm{1}_{ \{ y_{i,k}\leq  9 \frac{\sum_i y_{i,k}}{m} \} }$ scaled by $\sqrt{\sum_{i=1}^m \y_{i,k} / m }$. 
\end{algorithm}

We used $n=100$, $r=2$, $\a_{i,k} \iidsim \N(0,\I)$ (Table \ref{real_gauss}) and $\a_{i,k} \iidsim \mathcal{CN}(0,\I)$ (Table \ref{complex_gauss}) and varied $m$ and $q$. Here $\mathcal{CN}$ refers to a circularly symmetric complex Gaussian distribution. We show 100-time Monte Carlo averaged errors. The averaging is only over the measurement vectors.
As can be seen from both tables, LRPR-init significantly outperforms TWF-init and TWFproj-init when $m$ is small. 
The reason is that LRPR-init estimates $\X=\U \B$ by first estimating $\Span(\U)$  as the top $r$ eigenvectors of $\Y_U$; and $\Y_U$ averages the nearly $mq$ mutually independent matrices $\M_{i,k}:=\y_{i,k} \a_{i,k} \a_{i,k}{}'$. 
This is possible to do because $\E[\y_{i,k} \a_{i,k}\a_{i,k}{}'] = 2\U \b_k \b_k{}' \U' + \|\b_k\|^2 \I$. Thus even though the expected values are different for different $k$, all have span of top $r$ eigenvectors equal to $\Span(\U)$.
%
In TWF-init, the averaging over measurements of different columns is not exploited at all, and thus, unsurprisingly, it has the worst performance. 
In TWFproj-init, averaging over both $k$ and $i$ is exploited, but not simultaneously - the first step is TWF-init which averages only over $i$. The projection step can be interpreted as averaging over the $q$ rank-one matrices $\xhat_k \xhat_k{}'$ (where $\xhat_k = \xhat_k^{\mathrm{TWF,init}}$), followed by computing its top $r$ eigenvectors and projecting $\Xhat^{\mathrm{TWF,init}}$ onto their subspace. In situations where the TWF-init error itself is very large, the second step does not help much.

When the product $mq$ is large, the rank $\rhat$ is correctly estimated by LRPR-init (Algorithm \ref{algo_init}) either always or most of the time. We display a Monte Carlo estimate of the probability of $\rhat=r$ in the 2nd column.
In these cases, LRPR with $\rhat$ known versus $\rhat$ estimated both have similar errors (3rd and 4th columns).
Inspired by a reviewer's concern, we also evaluate LRPR-init with $\hat{r}$ deliberately set to a wrong value $2r$ in the 5th column. As can be seen, the error degradation is gradual even with a wrong rank estimate. 

Finally, Table \ref{real_gauss} shows errors of {\em LRPR-Same} in the last column. This refers to LRPR operating on measurements of the form $\y_{i,k}:=(\a_i{}'\x_k)^2$. Because it uses the same $\a_i$'s for all columns $\x_k$,
there are only $m$ (and not $mq$) mutually independent matrices to average over. Hence its errors are almost as large as those of TWF.


\begin{figure*}[t!]
\begin{subfigure}{.33\textwidth}
  \centering
  \includegraphics[width=1\linewidth,height=3.5cm]{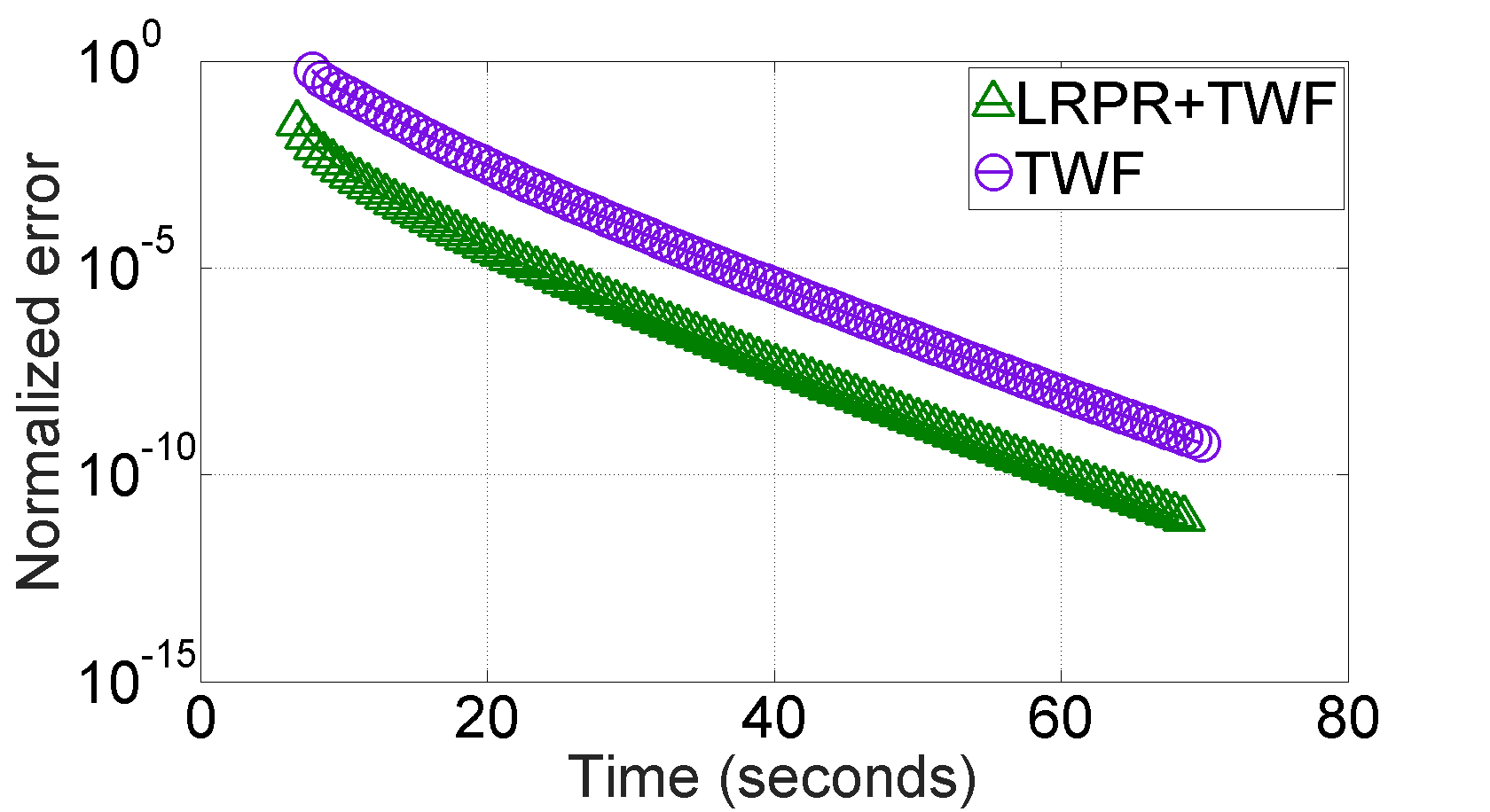}
  \caption{$m=8n$, $q=1000$ }
  \label{TWF_compare}
\end{subfigure}
\begin{subfigure}{.33\textwidth}
  \centering
  \includegraphics[width=1\linewidth,height=3.5cm]{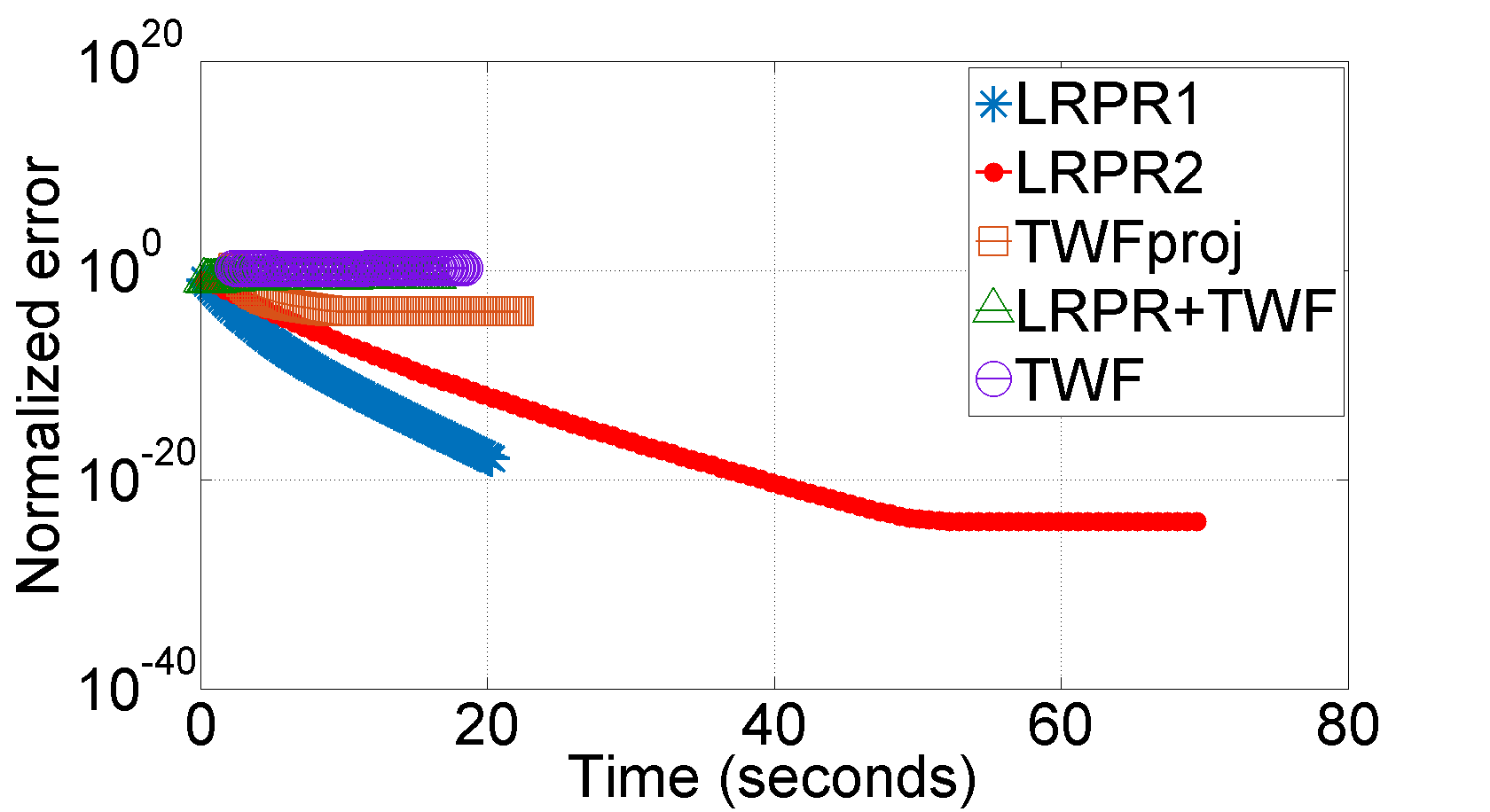}
  \caption{$m=0.8n$, $q=1000$}
  \label{all_compare}
\end{subfigure}
\begin{subfigure}{.33\textwidth}
  \centering
  \includegraphics[width=1\linewidth,height=3.5cm]{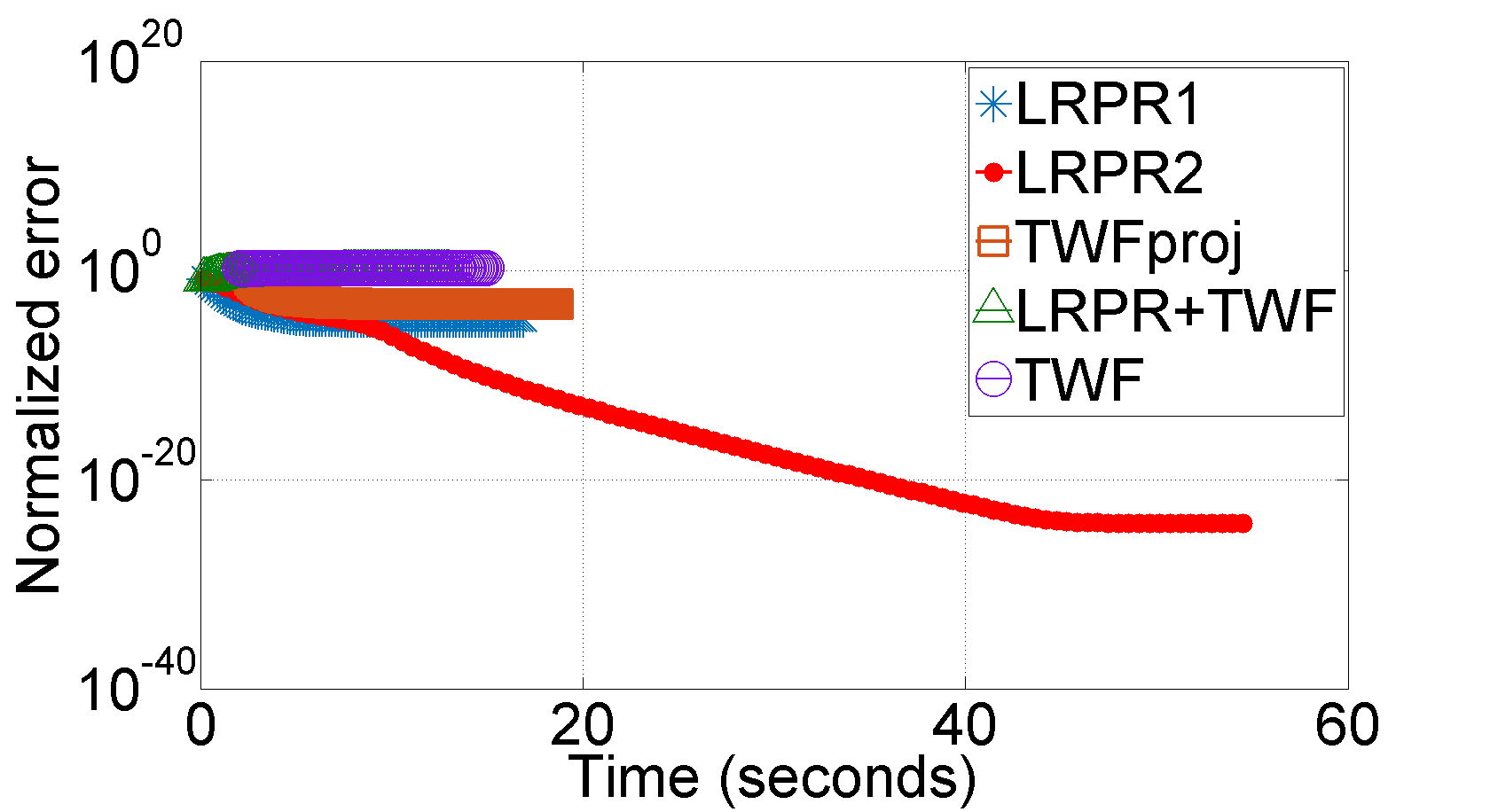} 
  \caption{$m=0.6n$, $q=1000$}
  \label{altmin_compare}
\end{subfigure}
\caption{\small\sl{Plot of reconstruction error, $\mathrm{NormErr}$, as a function of the computation time taken. We obtain each plot as follows. For each iteration $t=0,1,2,\dots,100$, we plot the error at iteration $t$ against the time taken until iteration $t$. This is why, for algorithms with lower per iteration cost, the plot ends earlier, e.g., in (b),  LRPR1 took only 20 seconds to complete 100 iterations and hence its plot ends at that time.
}}
\label{fig:fig}
\end{figure*}

{\bf Experiment 2. }
We evaluated LRPR-init (Algorithm \ref{algo_init}) in the noisy measurements case. We generated $\y_{i,k} = (\a_{i,k}{}' \x_k)^2 + \w_{i,k}$ where $\w_{i,k}$ were iid uniform between $-1$ and $1$ (so that the noise variance was $0.33$). The Monte Carlo estimate of $\E[|\a_{i,k}{}' \x_k|^2]$ was 0.67 leading to a signal-to-noise ratio of $2$.
The results are shown in the 6th columns of Tables \ref{real_gauss} and \ref{complex_gauss}. 
Observe that the error of LRPR-init even with noisy measurements is smaller than the errors of TWFproj-init and TWF-init with noise-free measurements.

{\bf Experiment 3. }
For various values of $m$, we evaluate the speed of convergence of the five complete algorithms -
basic TWF (Algorithm \ref{twf_iter_algo} initialized with Algorithm \ref{twf_init_algo}),
basic TWFproj (Algorithm \ref{twfproj_init_algo} initialized with  Algorithm \ref{twfproj_iter_algo}),
LRPR+TWF (Algorithm \ref{twf_iter_algo}),
LRPR1 (Algorithm \ref{twfproj_iter_algo}), and
LRPR2 (Algorithm \ref{amt}).
We define {\em ``converges" as $\mathrm{NormErr}$ below $10^{-10}$}. We generated data as in the noise-free complex Gaussian case described above with $n=100$, $r=2$ and $q=1000$.

In Fig. \ref{TWF_compare}, we compare the speed of error decay of TWF when initialized with either TWF-init (TWF) or with the proposed initialization, LRPR-init (LRPR+TWF). We used $m=8n$ (large enough $m$ for TWF iterations to converge). For $t=0,1,2,\dots,100$, we plot the error at the end of iteration $t$ on the y-axis and the time taken till the end of iteration $t$ on the x-axis ($t=0$ corresponds to initialization). As can be seen, LRPR-init takes longer to finish than TWF-init (the first `triangle' is to the right of the first circle). However, because LRPR-init results in much lower initialization error, LRPR+TWF needs much fewer iterations to ``converge", and, so the total time taken by it to ``converge" is also smaller.

If $m$ is reduced to $m=0.8n$ measurements, as can be seen from Fig. \ref{all_compare}, neither of TWF or LRPR+TWF converge. 
Basic TWFproj also does not converge and this is because its initialization error is larger (for reasons explained earlier).
However, both LRPR1 and LRPR2 converge. It is also apparent that LRPR1 is significantly faster than LRPR2. This is because its per iteration cost is lower.

If $m$ is reduced further to $m=0.6n$ (Fig. \ref{altmin_compare}), then LRPR1 does not converge whereas LRPR2 still does. This is because LRPR2 iterates directly exploit the split-up $\X=\U \B$ whereas LRPR1 iterates first implement a TWF iteration and then project the resulting matrix onto the space of rank $r$ matrices.

%% file: plane_expt_describe.tex
\section{Experiment details for Fig. \ref{planefig}}
We used real videos that are approximately low rank and CDP measurements of their images. Each image (arranged as a 1D vector) corresponds to one $\x_k$ and hence the entire video corresponds to the matrix $\X$.
We show results on a moving mouse video and on a moving airplane video (shown in Fig. \ref{planefig}).
We show two results with  ``low-rankified videos"\footnote{The original video data matrix $\X_{orig}$ was made exactly low rank by projecting it onto the space of rank-$r$ matrices where $r$ was chosen to retain 90\% of the singular values' energy.} and one result with the original airplane video. The airplane images were of size $n_1 \times n_2$ with $n_1=240$, $n_2 = 320$; the mouse images had $n_1=180$, $n_2 = 319$. Thus,  $n=n_1 n_2 = 76800$ and $n=57420$ respectively.
Mouse video had $q=90$ frames and airplane one had $q=105$ frames.

The CDP measurement model can be understood as follows \cite{twf}. First, note that it allows $m$ to only be an integer multiple of $n$; so let $m=n L$ for an integer $L$.
Let $\y_k$ denote the vector containing all measurements of $\x_k$. Then $\y_k = |\A_k{}' \x_k|^2$ where $\A_k = [ (\F\M_{k,1})' , ( \F\M_{k,2})', \dots ,  \F (\M_{k,L})']$; each $\M_{k,l}$ is a diagonal  $n \times n$ mask matrix with diagonal entries chosen uniformly at random from the set $\{1,-1, \sqrt{-1}, - \sqrt{-1}\}$, and $\F = \F_{1D,n_1} \otimes \F_{1D,n_2}$ where $\F_{1D,n}$ is the $n$-point discrete Fourier transform (DFT) matrix and $\otimes$ denotes Kronecker product. Thus, $(\F \x_k)$ is the vectorized version of the 2D-DFT of the image corresponding to $\x_k$

In this experiment, $n$ and $m$ are very large and hence the memory complexity is very large. 
Thus, the algorithm cannot be implemented using matrix-vector multiplies.
However, since the measurements are masked-Fourier, we can implement its ``operator" version as was also done in the TWF code \cite{twf}. All matrix-vector multiplies are replaced by ``operators" that use 2D fast Fourier transform (2D-FFT) or 2D-inverse-FFT (2D-IFFT) functions, preceded or followed by applying the measurement masks. This is a much faster and memory efficient implementation. Only the masks need to be stored. The EVD in the initialization step is implemented by a block-power method that uses 2D-FFT. The LS step is implemented using the operator-version of conjugate gradient LS (CGLS) taken from \url{http://web.stanford.edu/group/SOL/software/cgls/}. TWFproj and LRPR1 are implemented similarly.

For this experiment, we used 50 outer loop iterations in each algorithm. Also, 50 iterations of the block-power method were used. For LRPR2, 3 iterations of CGLS were used.
We display the $\mathrm{NormErr}$ for LRPR2, LRPR1, TWF (TWF-init+TWF) and TWFproj (TWFproj-init+TWFproj) in Table \ref{tab_real}. Execution times are again shown in parentheses. Three frames of the results corresponding to the last row of this table are shown in Fig. \ref{planefig} in Sec.~\ref{algos}.
As can be seen, LRPR2 has the smallest error in all cases. LRPR2 is also the slowest; it is at least $r$ times slower than TWF. But, TWF and TWFproj do not work when $m=nL$ is small: notice that the error is much more than one even for $L=3$.
LRPR1 is slower than TWF and TWFproj but is much faster than LRPR2. Notice also that, when LRPR2 error is more than 0.1, LRPR1 error is not too much larger than that of LRPR2; in the regime when LRPR2 error is below 0.001, LRPR1 error is 100-1000 times larger. Thus, if just a good approximate solution is needed, LRPR1 offers a better compromise between speed and performance with fewer measurements. If a very accurate solution is needed but speed is not a concern, LRPR2 is a better idea.


\begin{table} 
\centering{
\scalebox{0.96}{
\begin{tabular}{|c|c|c|c|}
\hline
{LRPR2} & {LRPR1} &  {TWF}&  {TWFproj}\\
\hline
\multicolumn{4}{|c|}{ {Mouse, Low-rankified video, $r=15,L=1$}}\\
\hline 
0.52 (981)& 0.65 (548)&NaN (54)& NaN (389)\\
\hline
\multicolumn{4}{|c|}{ {Mouse, Low-rankified video, $r=15, L=2$}}\\
\hline
8.0e-04 (18776)& 0.07 (905)& 2.2 (103)& 13 (394)\\
\hline
\multicolumn{4}{|c|}{{Plane, Low-rankified video, $r=6, L=2$}}\\
\hline
7.8e-10 (1036) & 6.9e-07 (574)& 2.2 (137) & 14 (327)\\
\hline
\multicolumn{4}{|c|}{{Plane, Original video, $r=6, L=2$}}\\
\hline
0.579 (1042)&0.583 (567) & 2.2 (134)& 14 (339)\\
\hline
\multicolumn{4}{|c|}{ {Plane, Original video, $r=25, L=3$}}\\
\hline
0.146 (13472)&0.150 (3451)& 2.0 (207)& 14  (950)\\
\hline
\end{tabular}
}
\caption{\small{Results for videos with CDP measurements: the table is displayed as NormErr (time in seconds). 
We use the symbol ``NaN" to indicate that the TWF or TWFproj code failed. This happens for the $L=1$ case (since $m=n$ is too few measurements for TWF). 
}}
\label{tab_real}
}
\vspace{-0.25in}
\end{table}